\documentclass[%
reprint,
amsmath,amssymb,
aps,
]{revtex4-1}

\usepackage{graphicx}
\usepackage{dcolumn}
\usepackage{bm}
\usepackage{bbm}
\usepackage{mathrsfs}

\usepackage[table]{xcolor}

\begin{document}
	
	\preprint{APS/123-QED}
	
	\title{\large\bf Probing Majorana bound states through \\ an inhomogeneous Andreev double dot interferometer}
	
	
	
	\author{S.\,V.\, Aksenov}%
	\email{asv86@iph.krasn.ru}
	
	\affiliation{%
		Kirensky Institute of Physics, Federal Research Center KSC SB RAS, 660036 Krasnoyarsk, Russia}
	
	\date{\today}
	
	\begin{abstract}

In modern experiments with hybrid superconducting (SC)/semiconducting nanowires the presence of zero-energy Andreev bound states (ABSs), characterized by a partial overlap of the Majorana wave functions, is a common problem that significantly complicates the detection of a genuine Majorana bound state (MBS). In this article, taking into account spatial inhomogeneity of experimentally investigated nanowire samples, we study interference transport features of a curved heterostructure in which two normal wires (or arms) are separated by a superconducting wire. Since Andreev reflection on the two N/S interfaces with smoothly changing electrostatic and SC pairing potentials results in the emergence of bound states, the low-energy interference transport is described in the framework of the model of two noninteracting Andreev levels or the Andreev double quantum dot. A set of limiting cases is analyzed allowing us to highlight the interference properties that are unique for the different types of ABSs, such as bulk ABS, inhomogeneous ABS and MBS. In particular, considerable attention is paid to the features of the Aharonov-Bohm (AB) effect. It is shown that the response of each state can be recognized analyzing both AB period and extrema positions of the conductance oscillations which take place without any fine tuning of the system parameters. 

\begin{description}
	\item[PACS number(s)]
	71.10.Pm, 
	74.78.Na, 
	74.45.+c, 
\end{description}

\end{abstract}	
		
\maketitle


\section{\label{sec1}Introduction}

Andreev reflection is a fundamental process defining the properties of normal metal/superconductor (N/S) heterostructure \cite{andreev-64}. It leads to the emergence of subgap Andreev bound states (ABSs) localized in the normal part. These Bogolyubov excitations play an essential role in current carrying phenomena in S/N/S junctions and superconducting (SC) quantum point contacts \cite{kulik-70,beenakker-91} which are the building blocks for the applications in SC electronics, quantum computing and simulations \cite{wendin-17,houck-12}. Recently, a lot of attention has been paid to the ABS spatial extension and the hybridization of these states. For example, these phenomena are crucial for the realization of Andreev molecular states \cite{su-17,pillet-19,kornich-19}.

Another area where the spatial structure of the ABS wave function has become an important factor is topological superconductivity, which is currently undergoing rapid development \cite{choy-11,sau-12,banerjee-22}. The popularity of this direction is largely due to the unique properties of a special type of ABS, namely, the Majorana bound state (MBS) \cite{elliott-15,valkov-22}. This subgap excitation consists of two zero-energy Majorana components (MCs) whose wave functions are separated in space. Possessing spatial nonlocality, such states may turn out to be a promising basic element for the implementation of quantum calculations, which are more resistant to incoherent scattering processes and, as a result, to information loss \cite{kitaev-03,nayak-08}. 

In the problem of MBS detection much attention is focused on the tunneling spectroscopy of hybrid semiconducting/superconducting nanowires \cite{oreg-10,lutchyn-10}. The existing experimental data, in particular, the observation of the quantized zero-bias peak of conductance \cite{mourik-12}, do not allow us to unambiguously determine the presence of MBSs and realization of the topological phase transition \cite{zhang-21,yu-21}, stimulating further development of the transport theory in systems with topological superconductors. 

As a result, it was found that the aforementioned zero-bias peak can be induced by ABSs of nontopological nature \cite{kells-12b,prada-12,prada-20}. The part of trivial ABSs is related to the presence of a quantum-dot region typically located at the nanowire edge (or both of them). In the experiments it can be formed due to the Schottky barrier between the lead and SC-covered segment \cite{mourik-12,deng-16,nichele-17}. It was demonstrated that for such a system in the trivial phase the ABSs emerge due to a smooth change of chemical potential and SC gap \cite{moore-18,pan-20b,hess-21} or when the resonant conditions for a spin-orbit coupling strength are satisfied \cite{reeg-18}. Note that depending on the specific properties of the inhomogeneity at the N/S interface (e.g., the degree of smoothness, quantum-dot area length, MC localization length), the overlap of two MC wave functions can be comparable with the one in the topological phase. As a result, nonlocal trivial ABSs occur which are also called partially separated ABSs, pseudo- or quasi-MBSs \cite{moore-18b,penaranda-18,fleckenstein-18}. Another mechanism that causes the ABSs is a random disorder \cite{bagrets-12,pan-20c,pan-21}. Essentially, the energies of all these states can be pinned close to zero value in an extended range of the magnetic fields and gate voltages that mimics the MBS response in the transport measurements and substantially complicates interpretation of the experimental data \cite{zhang-21,yu-21,zhang-19,song-21,wang-22,aghaee-22}.

\begin{figure}[tb]
	\includegraphics[width=0.45\textwidth]{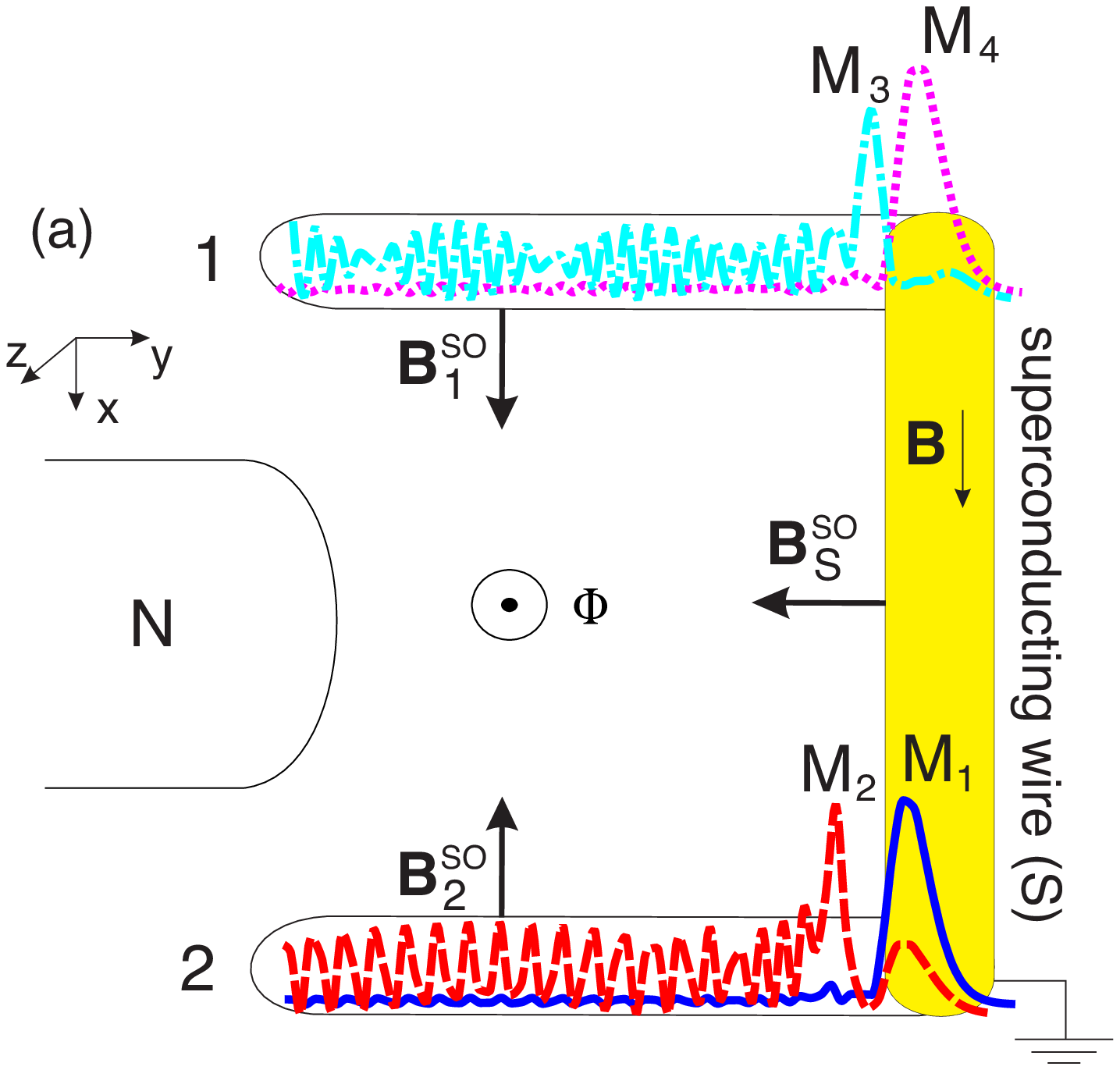}
	\includegraphics[width=0.47\textwidth]{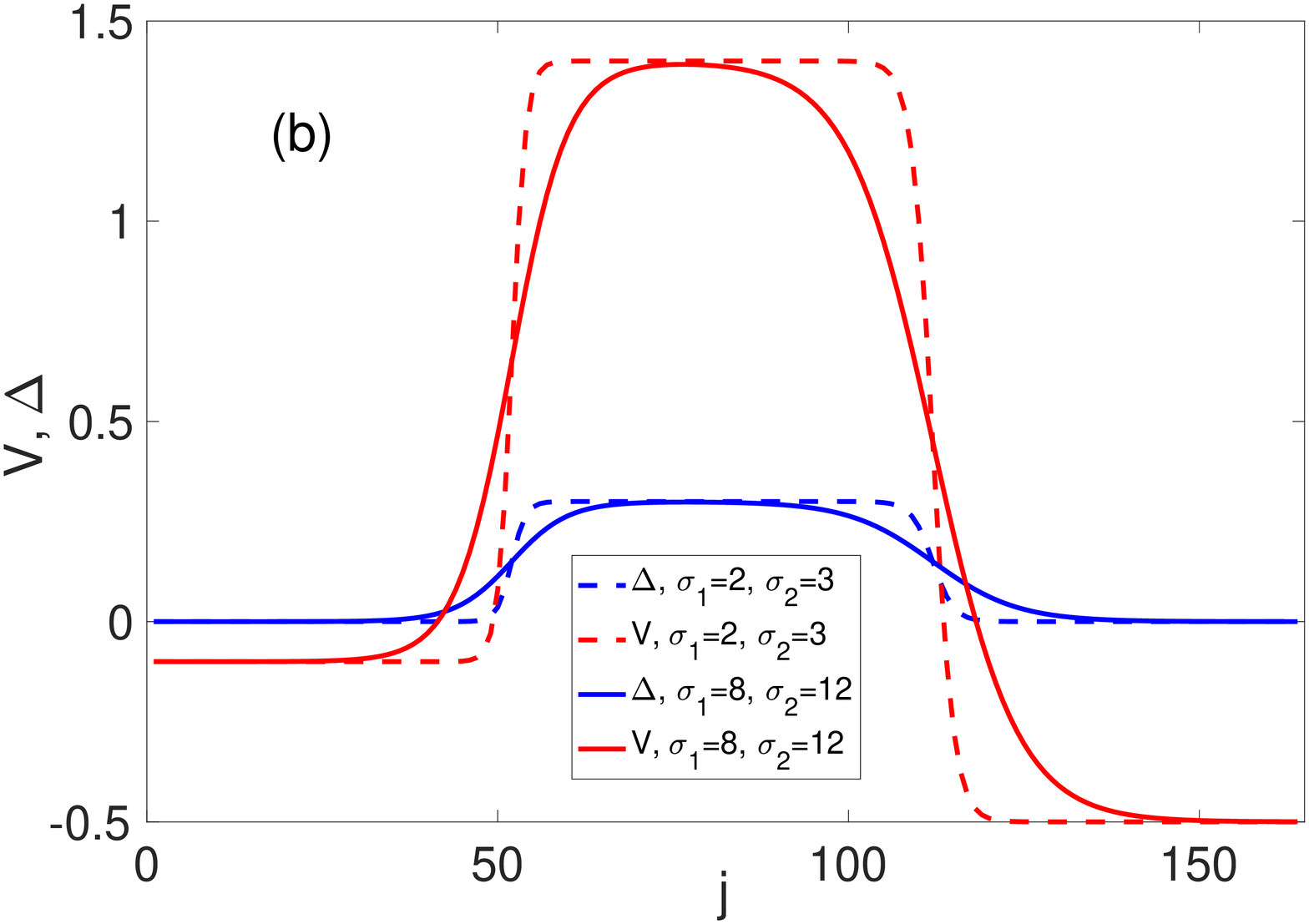}
	\caption{\label{1} (a) $\Pi$-shaped device with nonuniform profiles of the electrostatic, $V\left(j\right)$, and superconducting pairing, $\Delta\left(j\right)$, potentials. In general, their smooth dependence on a spatial variable at the N/S interfaces can induce two trivial Andreev bound states with zero energy, one each on half of the device. The probability densities of the Majorana components, $M_{1-4}$, composing these excitations are shown schematically by curves. An in-plane magnetic field, $\mathbf{B}$, is applied locally to a superconducting part of the device (S) and oriented parallel to it. A magnetic flux piercing the device plane induces the Aharonov-Bohm phase for the carriers tunneling from a normal contact (N) to the top and bottom arms of the device (denoted by '1' and '2' indices). The Rashba spin-orbit field, $\mathbf{B}^{SO}_{1,2,S}$, is present in all three sections of the device. (b) Profiles of the electrostatic and superconducting pairing potentials for 'steep-steep' (dashed curves) and 'gentle-gentle' (solid curves) cases. Parameters: $\varepsilon_{1}=-0.1$, $\varepsilon_{2}=-0.5$, $\varepsilon_{S}=1.4$, $\mu=0$, $\Delta_{0}=0.3$, $N_{1,2}=52$, $N_{S}=60$.}
\end{figure}

Here we study a one-dimensional interference heterostructure consisting of two normal wires (or arms) separated by an SC segment and coupled with a normal contact as displayed in Fig. \ref{1}a. It is assumed that the electrostatic and SC pairing potentials change smoothly at the N1/S and S/N2 interfaces. The corresponding profiles of $V\left(j\right)$ and $\Delta\left(j\right)$ are plotted in Fig. \ref{1}b. As was mentioned above such a behavior induces the trivial zero-energy ABSs in both arms (or inhomogeneous ABSs). It allows us to explain the low-energy physics employing a model of an inhomogeneous Andreev double quantum dot. The nonuniformity manifests itself in the asymmetric coupling of the contact with the Majorana components constituting the separate ABS that can be probed by the measurement of linear-response conductance. 

Due to the bent shape it becomes possible to analyze the features of local interference (the definition "local" means that there is only one normal-metal contact) caused by the ABSs of different type, i.e. the usual bulk states, inhomogeneous ABSs (including quasi-MBSs) and topological MBSs, and to distinguish between them. In other words, the proposed setup gives an opportunity to test not only the appearance of the zero-bias conductance peak but, additionally, the Aharonov-Bohm (AB) effect avoiding the utilization of nonlocal techniques, for example, measurement of transconductance \cite{rosdahl-18} or current correlations in different normal leads \cite{haim-15}. It is assumed that the latter are necessary to confirm the nontrivial nature of the subgap states. 

It is worth to note that the found transport properties of the inhomogeneous Andreev double quantum dot are in good agreement with the numerical results obtained for the microscopic model of the $\Pi$-shaped device depicted in Fig. \ref{1}. Thus, one can expect that the predicted effects do not qualitatively depend on the specific form of the setup. Taking it into account, there are several options for a potential verification experiment. Keeping in mind recent studies of AB interferometers that have been made using InSb nanowires \cite{gazibegovic-17,borsoi-20} or InAs 2DEG \cite{whiticar-20}, a triangular-shaped device can be investigated in which the SC wire is the base of the triangle, while its sides are straight or smoothly curved normal segments, respectively.

The rest of article is organized as follows: in Sec. \ref{sec2} we describe the Hamiltonian of the system. In Sec. \ref{sec3} we derive a linear-response conductance formula for the Andreev double dot taking into account the generally asymmetric couplings of the MCs with the reservoir caused by the presence of the smooth inhomogeneities. The obtained expression is a spinful generalization of the result obtained in \cite{vuik-19} for the case of spin-independent transport in the single arm. In Sec. \ref{sec4} we analyze the spectral and transport properties which are found numerically for a device microscopic Hamiltonian and are in good agreement with the effective model predictions. We conclude in Sec. \ref{sec5} with a summary.

\section{\label{sec2} Model description}

Let us define the device tight-binding Hamiltonian in terms of the Gorkov-Nambu spinors, $\hat{\psi}_{mi}^{T}=\left(a_{mi\uparrow}~~a_{mi\downarrow}^{+}~~a_{mi\downarrow}~~a_{mi\uparrow}^{+}\right)$, with the components that are annihilation and creation operators of the electron on the $i$th site in the $m$th section ($m=1,2,S$), i.e.
\begin{equation} \label{HD}
\hat{H}_{D}=\sum\limits_{n=1,2}\left[\hat{H}_{n}+\hat{H}_{Sn}\right]+\hat{H}_S,
\end{equation}
where
\begin{eqnarray} 
&&\hat{H}_{n} =\frac{1}{2}\sum\limits_{i=1}^{N_{n}}\hat{\psi}^+_{ni}\left[\left(V_{ni}-\mu\right)\hat{\tau}_{z}-h_{ni}\hat{\sigma}_{x}-\Delta_{ni}\hat{\tau}_{x}\right]\hat{\psi}_{ni}-\nonumber\\
&&~~~~~~-\frac{1}{2}\sum\limits_{i=1}^{N_{n}-1}\left[\hat{\psi}^+_{ni}\left(t-i\alpha_{n}\hat{\sigma}_{x}\right)\hat{\tau}_{z}\hat{\psi}_{n,i+1}+H.c.\right],\label{Hn}\\
&&\hat{H}_{S} =\frac{1}{2}\sum\limits_{i=1}^{N_{S}}\hat{\psi}^+_{Si}\left[\left(V_{Si}-\mu\right)\hat{\tau}_{z}-h_{Si}\hat{\sigma}_{x}-\Delta_{Si}\hat{\tau}_{x}\right]\hat{\psi}_{Si}-\nonumber\\
&&~~~~~~-\frac{1}{2}\sum\limits_{i=1}^{N_{S}-1}\left[\hat{\psi}^+_{Si}\left(t-i\alpha_{S}\hat{\sigma}_{y}\right)\hat{\tau}_{z}\hat{\psi}_{S,i+1}+H.c.\right],\label{HS}\\
&&\hat{H}_{S1}=-\frac{1}{2}\hat{\psi}^+_{1N_{1}}\left(t-i\alpha_{1}\hat{\sigma}_{x}\right)\hat{\tau}_{z}\hat{\psi}_{S1}+H.c.,\nonumber\\
&&\hat{H}_{S2}=-\frac{1}{2}\hat{\psi}^+_{SN_{S}}\left(t-i\alpha_{2}\hat{\sigma}_{x}\right)\hat{\tau}_{z}\hat{\psi}_{21}+H.c..\label{HSn}
\end{eqnarray}
The first two terms in \eqref{HD} are Hamiltonians of the 1st and 2nd predominantly normal wires including $N_{1}$ and $N_{2}$ sites, respectively, and serving as the top and bottom arms of the interferometer (see Fig. \ref{1}a). In turn, the last summand in \eqref{HD} is a Hamiltonian of the predominantly SC wire containing $N_{S}$ sites and being, in fact, one of the contacts. The remaining terms in \eqref{HD} describe an interaction between the three sections. 

The specification 'predominantly' here indicates the spatially inhomogeneous character of the total SC pairing potential, $\Delta\left(j\right)=\Delta_{1i}\cup\Delta_{Si}\cup\Delta_{2i}$ ($j=1,...,N$, $N=N_{1}+N_{S}+N_{2}$), that changes smoothly at the N1/S and S/N2 interfaces. Such a feature can be attributed, for example, to imperfect covering of the semiconducting core with the SC material leading to the appearance of its tails in the normal wire segments. The corresponding functional dependence is set by the following standard expression \cite{penaranda-18,fleckenstein-18,moore-18b,hess-21}:
\begin{equation} \label{Delj}
\Delta\left(j\right)=\frac{\Delta_{0}}{2}\left[\tanh\left(\frac{j-N_{1}}{\sigma_{1}}\right)-\tanh\left(\frac{j-N_{1}-N_{S}}{\sigma_{2}}\right)\right],
\end{equation}
where $\Delta_{0}$ is a bare SC gap; $\sigma_{1,2}$ are parameters defining a smoothness degree of the profile at the N1/S and S/N2 interfaces, respectively. These parameters are assumed to be the same for the electrostatic potential profile, $V\left(j\right)=V_{1i}\cup V_{Si}\cup V_{2i}$, i.e.
\begin{eqnarray} \label{Vj}
&&V\left(j\right)=\frac{\varepsilon_{1}+\varepsilon_{2}}{2}+\frac{\varepsilon_{S}-\varepsilon_{1}}{2}\tanh\left(\frac{j-N_{1}}{\sigma_{1}}\right)-\\
&&~~~~~~~~~~~~~~~~~~~~~~~~~~~~-\frac{\varepsilon_{S}-\varepsilon_{2}}{2}\tanh\left(\frac{j-N_{1}-N_{S}}{\sigma_{2}}\right)\nonumber,
\end{eqnarray}
where $\varepsilon_{1,2,S}$ are the on-site energies in the three subsystems. As was already mentioned in Sec. \ref{sec1}, the smoothness of the $V\left(j\right)$ profile can be caused by the presence of the Schottky barriers and gate electrodes. During the numerical calculations of spectral and transport properties of the interferometer we will consider two limiting cases of steep and gentle slopes (or their combination for the opposite interfaces) for the $\Delta\left(j\right)$ and $V\left(j\right)$ profiles depicted in Fig. \ref{1}b by the dashed and solid curves, respectively.

Next, the inhomogeneities at the two boundaries related to the Zeeman energy and spin-orbit interaction are supposed to have a step-like character. In particular, an in-plane magnetic field (real or effective \cite{vaitiekenas-21}) is assumed to be applied locally to the S-part, i.e. $h_{1,2;i}=0$ and $h_{Si}=h$. The effective Rashba field $\mathbf{B}_{m}^{SO}$ ($m=1,2,S$) with the amplitude proportional to $\alpha_{m}$ rotates $90$ degrees when moving from the one section to the other (see Fig. \ref{1}a). Finally, in \eqref{Hn}-\eqref{HSn} $\mu$ and $t$ are a chemical potential and hopping parameter. The Pauli matrices $\hat{\sigma}$ and $\hat{\tau}$ act in the spin and particle-hole subspaces, respectively. 

As displayed in Fig. \ref{1}a the interference device is coupled with the normal contact that is modeled by a standard single-band Hamiltonian, $\hat{H}_{N}=\frac{1}{2}\sum_{k}\hat{\psi}^+_{k}\left(\varepsilon_{k}-eU/2-\mu\right)\hat{\tau}_{z}\hat{\psi}_{k}$, where $\hat{\psi}_{k}^{T}=\left(c_{k\uparrow}~~c_{k\downarrow}^{+}~~c_{k\downarrow}~~c_{k\uparrow}^{+}\right)$. The interaction between the normal contact biased with a voltage $U/2$ and the directly grounded superconductor is implemented via the two arms and is described by the following tunnel Hamiltonian:
\begin{equation} 
\hat{H}_{T}=-\hat{\psi}^+_{k}\hat{\tau}_{z}\left(\hat{t}_{1}\hat{\Phi}\hat{\psi}_{11}+\hat{t}_{2}\hat{\Phi}^{+}\hat{\psi}_{2N_{2}}\right)+H.c.,
\end{equation}
where $\hat{t}_{1,2}$ are matrices containing the tunneling coefficients which are, in general, spin-dependent and complex, i.e. $\hat{t}_{1\left(2\right)}= diag\left(t_{1\left(2\right)\uparrow},t_{1\left(2\right)\downarrow}^{*},t_{1\left(2\right)\downarrow},t_{1\left(2\right)\uparrow}^{*}\right)$; $\hat{\Phi}=diag\left(e^{i\frac{\phi}{2}},e^{- i\frac{\phi}{2}},e^{ i\frac{\phi}{2}},e^{- i\frac{\phi}{2}}\right)$ includes the AB phase $\phi=2\pi\Phi/\Phi_{0}$ due to a magnetic flux $\Phi$ penetrating the device; $\Phi_{0}=h/e$ is the flux quantum.

\section{\label{sec3} Spinful zero-bias conductance of inhomogeneous Andreev double quantum dot}

Since, in general, there are two interfaces, N1/S and S/N2, where the electrostatic and SC pairing potentials smoothly change, one has to expect that the trivial zero-energy ABSs can emerge in each N-section and the adjacent edge of the S-segment \cite{prada-20}. Therefore, aiming at an explanation of low-energy transport features of such an Andreev-double-dot structure we derive the corresponding expression for linear-response conductance. In this regime only local Andreev reflection contributes to the current between the normal and SC contacts. Then, the conductance in terms of the scattering matrix, $\hat{S}$, can be written as
\begin{equation} \label{GS}
G_{\omega}=2G_{0} Tr\left[\hat{S}_{eh}^{+}\hat{S}_{eh}\right],~~~\hat{S}=\left({\begin{array}{*{2}{c}}
	\hat{S}_{ee}&\hat{S}_{eh}\\
	\hat{S}_{he}&\hat{S}_{hh}
	\end{array}} \right).
\end{equation}
Using the Mahaux-Weidenm\"{u}ller formula \cite{mahaux-68,mahaux-69,fisher-81} the S-matrix is expressed via the operator of single-particle retarded Green's function for the open system (i.e. the device interacting with the reservoir), $\hat{g}^{r}$, \cite{datta-95}
\begin{equation} \label{MWformula}
\hat{S}=\mathbbm{1}-2\pi i \hat{W} \hat{g}^{r} \hat{W}^{+}, ~\hat{g}^{r}=\left(\omega\cdot\mathbbm{1}-\hat{H}+i\pi \hat{W}^{+}\hat{W}\right)^{-1},
\end{equation}
where $\hat{W}=\hat{t}'_{1}\hat{\Phi}\hat{I}_{1}+\hat{t}'_{2}\hat{\Phi}^{+}\hat{I}_{N}$ is a tunneling matrix; $\hat{t}'_{1\left(2\right)}=\sqrt{\rho}\cdot \hat{t}_{1,2}$; $\rho$ is a density of states of the normal contact which is assumed to be constant;  $\mathbbm{1}$ is a $4$-by-$4$ unity matrix; $\hat{I}_{1\left(N\right)}$ is a $4$-by-$4N$ matrix where only the $1$st ($N$th) block is nonzero and equal to $\mathbbm{1}$. 

As noted above we are interested in the transport mediated by the two eigenstates of $\hat{H}_{D}$ localized in the opposite halves of the structure. In order to make the underlying physics related to the spatial inhomogeneity more clear it turns out to be useful to pass to the basis of Majorana wave functions,
\begin{eqnarray} \label{psiM} &&\psi_{1,3}^{M}=\frac{i}{\sqrt{2}}\left(\psi_{n}^{h}-\psi_{n}^{e}\right)=\frac{i}{\sqrt{2}}\sum_{j}\hat{w}_{jn}^{T}\hat{\tau}_{z}\hat{a}_{j}, \\ &&\psi_{2,4}^{M}=\frac{1}{\sqrt{2}}\left(\psi_{n}^{e}+\psi_{n}^{h}\right)=\frac{1}{\sqrt{2}}\sum_{j}\hat{z}_{jn}^{T}\hat{a}_{j},~n=a,b. \nonumber
\end{eqnarray}
Here $\psi_{a,b}^{e,h}$ are the electron- and hole-like Bogolyubov excitations with energies $E_{a,b}=\pm A,~\pm B$, respectively. In \eqref{psiM} the Gorkov-Nambu representation is used, namely,
\begin{eqnarray} \label{wza_Nambu} &&\hat{w}_{jn}^{T}=\left[v_{jn\uparrow}^{*}-u_{jn\uparrow}~v_{jn\downarrow}-u_{jn\downarrow}^{*}~v_{jn\downarrow}^{*}-u_{jn\downarrow}~v_{jn\uparrow}-u_{jn\uparrow}^{*}\right], \nonumber\\ &&\hat{z}_{jn}^{T}=\left[v_{jn\uparrow}^{*}+u_{jn\uparrow}~v_{jn\downarrow}+u_{jn\downarrow}^{*}~v_{jn\downarrow}^{*}+u_{jn\downarrow}~v_{jn\uparrow}+u_{jn\uparrow}^{*}\right], \nonumber\\
&&\hat{a}_{j}^{T}=\left[a_{j\uparrow}~a_{j\downarrow}^{+}~a_{j\downarrow}~a_{j\uparrow}^{+}\right].
\end{eqnarray}
The spatial dependence of the probability densities related to the wave functions \eqref{psiM}, i.e. $M_{1\left(3\right)}\left(j\right)=\hat{w}_{ja\left(b\right)}^{+}\hat{w}_{ja\left(b\right)}$ and $M_{2\left(4\right)}\left(j\right)=\hat{z}_{ja\left(b\right)}^{+}\hat{z}_{ja\left(b\right)}$, is schematically displayed in Fig. \ref{1}a.  

The projection onto the Hilbert subspace spanned by the four MCs that is executed by the operator $\hat{P}_{M}=\left[\psi_{1}^{M}~\psi_{2}^{M}~\psi_{3}^{M}~\psi_{4}^{M}\right]$ gives the following Hamiltonian and tunneling matrix:  
\begin{eqnarray} \label{HM}
&&\hat{H}_{M}=\left({\begin{array}{*{4}{c}}
	0&iA&0&0\\
	-iA&0&0&0\\
	0&0&0&iB\\
	0&0&-iB&0
	\end{array}} \right),~\\ 
&&\hat{W}_{M}=\frac{\hat{t}_{1}\hat{\Phi}}{\sqrt{2}}\left[i\hat{\tau}_{z}\hat{w}_{11}~\hat{z}_{11}~\hat{O}\right]+\frac{\hat{t}_{2}\hat{\Phi}^{+}}{\sqrt{2}}\left[\hat{O}~i\hat{\tau}_{z}\hat{w}_{N2}~\hat{z}_{N2}\right]=\nonumber\\
&&~~~~~~~~~~=\left[i\hat{\tau}_{z}\hat{\Phi}\hat{\tau}_{1}~~~\hat{\Phi}\hat{\tau}_{2}~~~i\hat{\tau}_{z}\hat{\Phi}^{+}\hat{\tau}_{3}~~~\hat{\Phi}^{+}\hat{\tau}_{4}\right], \label{WM}
\end{eqnarray}
In the first line of \eqref{WM}, the zero blocks $\hat{O}$ are the consequence of the localization of $\psi_{1,2}^{M}$ and $\psi_{3,4}^{M}$ in the opposite halves of the structure (see Fig. \ref{1}a). Next, to simplify the subsequent derivation of the conductance formula it is supposed that the $t_{1,2\sigma}$ phase adjustment allows us to consider the intensities of interaction between $\sigma$-spin transport channel of the reservoir and $i$th MC, $\tau_{i\sigma}$, as real numbers \cite{nilsson-08,vuik-19}. 

\begin{widetext}

\subsection{\label{sec3.1} Features of Aharonov-Bohm oscillations}

Substitution of $\hat{H}_{M}$ and $\hat{W}_{M}$ into \eqref{MWformula} and following calculation of the linear-response conductance, $G_{\omega=0}\equiv G$, results in
\begin{equation} \label{G_triv}
G_{triv}=4G_{0}\frac{A^2\Delta\overline{\Gamma}_{34}^{2}+B^2\Delta\overline{\Gamma}_{12}^{2}+2AB\Delta\overline{\Gamma}_{12}\Delta\overline{\Gamma}_{34}\cos2\phi}{\Theta_{triv}+\left(\Delta\overline{\Gamma}_{13}\Delta\overline{\Gamma}_{24}-\Delta\overline{\Gamma}_{12}\Delta\overline{\Gamma}_{34}\sin^{2}\phi\right)^2},
\end{equation}
where
\begin{eqnarray} \label{ThGm_triv}
&&\Theta_{triv}=\left(A\Gamma_{34}+B\Gamma_{12}\right)^{2}+\left(A\Delta\overline{\Gamma}_{34}+B\Delta\overline{\Gamma}_{12}\right)^{2}+AB\left(AB-2\Delta\Gamma_{12}\Delta\Gamma_{34}-2\Delta\overline{\Gamma}_{13}\Delta\overline{\Gamma}_{24}-2\Delta\overline{\Gamma}_{12}\Delta\overline{\Gamma}_{34}\sin^{2}\phi\right)\nonumber\\
&&\Gamma_{ij}=\sum\limits_{\sigma}\Gamma_{i\sigma,j\sigma},~\overline{\Gamma}_{ij}=\sum\limits_{\sigma}\Gamma_{i\sigma,j\bar{\sigma}},~\Delta\Gamma_{ij}=\sum\limits_{\sigma}\sigma\Gamma_{i\sigma,j\sigma},~\Delta\overline{\Gamma}_{ij}=\sum\limits_{\sigma}\sigma\Gamma_{i\sigma,j\bar{\sigma}}
\end{eqnarray}
and the broadening parameters related to the MCs are defined as $\Gamma_{i\sigma,j\sigma'}=2\pi\tau_{i\sigma}\tau_{j\sigma'}$ ($i,j=1,...,4$). Note that the AB oscillations are $\pi$-periodic and the AB effect is present unless $\Delta\overline{\Gamma}_{12}\Delta\overline{\Gamma}_{34}=0$. 

If the S-segment is in the topologically nontrivial phase then the 2nd MC moves to the opposite N-section of the device. Taking it into account, we obtain the following expression for the zero-bias conductance:
\begin{equation} \label{G_topo}
G_{topo}=4G_{0}\frac{A^2\Delta\overline{\Gamma}_{34}^{2}+B^2\Delta\overline{\Gamma}_{12}^{2}+2AB\Delta\overline{\Gamma}_{12}\Delta\overline{\Gamma}_{34}\cos\phi}{\Theta_{topo}+B^{2}\Delta\overline{\Gamma}_{12}^{2}\sin^{2}\phi},
\end{equation}
where
\begin{equation} \label{Th_topo}
\Theta_{topo}=\left[A\Gamma_{34}+B\Gamma_{12}\cos\phi\right]^{2}+\left[A\Delta\overline{\Gamma}_{34}+B\Delta\overline{\Gamma}_{12}\cos\phi\right]^{2}+\left[AB-\Delta\overline{\Gamma}_{13}\Delta\overline{\Gamma}_{24}\cos\phi\right]^{2}.
\end{equation}
Thus, in general, the $\phi$-dependence of conductance is $2\pi$-periodic. 

When the hybridization of the first and second MCs is negligible, i.e. $A\to0$, 
\begin{equation} \label{G_triv0} 
G_{triv}^{\left(0\right)}=4G_{0}\frac{B^2\Delta\overline{\Gamma}_{12}^{2}}{B^2\left[\Delta\overline{\Gamma}_{12}^{2}+\Gamma_{12}^{2}\right]+\left[\Delta\overline{\Gamma}_{13}\Delta\overline{\Gamma}_{24}-\Delta\overline{\Gamma}_{12}\Delta\overline{\Gamma}_{34}\sin^{2}\phi\right]^{2}},
\end{equation}
\begin{equation} \label{G_topo0} 
G_{topo}^{\left(0\right)}=4G_{0}\frac{B^2\Delta\overline{\Gamma}_{12}^{2}}{B^2\left[\Delta\overline{\Gamma}_{12}^{2}+\Gamma_{12}^{2}\cos^{2}\phi\right]+\Delta\overline{\Gamma}_{13}^{2}\Delta\overline{\Gamma}_{24}^{2}\cos^{2}\phi}.
\end{equation}
Therefore, the nontrivial conductance becomes $\pi$-periodic under the $\phi$ change. In spite of that one can still distinguish between the different states if $\phi=\pi\left(n+1/2\right)$, $n\in\mathbb{Z}$, as the trivial conductance reaches $4G_{0}$-quantized maximum only at a certain ratio between the tunneling amplitudes $\tau_{i\sigma}$ whereas $G_{topo}^{\left(0\right)}$ equals $4G_{0}$ for any $B,~\tau_{i\sigma}$ \cite{aksenov-22}. 

In general, the different periodicities can imply that both inhomogeneous ABSs are needed to generate the AB oscillations in the trivial phase while the single nonlocal excitation (i.e. the MBS) is sufficient in the nontrivial phase. It is confirmed in the regime $B \gg A,~\Gamma_{i\sigma,j\sigma'}$ where $G_{triv}\neq G_{triv}\left(\phi\right)$ and $G_{topo}=G_{topo}\left(\phi\right)$. Moreover, to obtain the same result, one can simply set the interaction parameters related to the second ABS to zero, $\tau_{3,4\sigma}=0$. Since the condition $B \gg \Gamma_{i\sigma,j\sigma'}$ can be achieved in practice employing gate electrodes, the absence/presence of the AB oscillations itself is another marker allowing us to distinguish between the inhomogeneous ABSs and MBSs.

Despite this, for a more complete analysis, it is necessary to additionally consider the interference transport mediated by a typical low-energy ABS with a spatial distribution throughout the device (i.e. a bulk ABS). It can emerge, for example, if the topological gap collapses at the high magnetic fields \cite{aksenov-22} or due to the presence of random disorder in the trivial phase \cite{pan-20b}. In the framework of the above-formulated model, both MCs of such an excitation interact with the normal contact via both arms. Neglecting for simplicity the contribution from the second Bogolyubov state (that is guaranteed if $A \ll B$) and denoting the coupling constants of the first (second) MC as $\tau_{1,3\sigma}$ ($\tau_{2,4\sigma}$), the conductance can be expressed as
\begin{equation} \label{G_bulk}
G_{bulk}=4G_{0}\frac{a_{1}\cos^{2}\phi+b_{1}\cos\phi+c_{1}}{a_{2}\cos^{2}\phi+b_{2}\cos\phi+c_{2}},
\end{equation}
where $a_{1}=4\Delta\overline{\Gamma}_{12}^{2}\Delta\overline{\Gamma}_{34}^{2}$, $b_{1}=2\left(\Delta\overline{\Gamma}_{12}+\Delta\overline{\Gamma}_{34}\right)\left(\Delta\overline{\Gamma}_{14}-\Delta\overline{\Gamma}_{23}\right)$, $c_{1}=\left(\Delta\overline{\Gamma}_{12}-\Delta\overline{\Gamma}_{34}\right)^{2}+\left(\Delta\overline{\Gamma}_{14}-\Delta\overline{\Gamma}_{23}\right)^{2}$, $a_{2}=\left(\Gamma_{14}-\Gamma_{23}\right)^{2}+4\Gamma_{13}\Gamma_{24}$, $b_{2}=b_{1}+2\left(\Gamma_{12}+\Gamma_{34}\right)\left(\Gamma_{14}-\Gamma_{23}\right)$, $c_{2}=c_{1}+\left(\Gamma_{12}+\Gamma_{34} \right)^{2}+A^{2}$. Note that $G_{bulk}\left(\tau_{3,4\sigma}=0\right)=G_{triv}\left(B\to\infty\right)$ and $G_{bulk}\left(\tau_{2,3\sigma}=0\right)=G_{topo}\left(B\to\infty\right)$.

In general, the AB oscillation period in \eqref{G_bulk} is $2\pi$ implying that this property alone is not sufficient for unambiguous characterization of the various states utilizing solely the interference picture. To solve this problem we suggest to supplement the study with the analysis of the conductance extrema at $\phi=\pi n/2$, $n\in\mathbb{Z}$ which are parameter independent. The final results for the most general situation, when only robust properties are considered, are presented in the following table:

\begin{table}[!htb]
	\begin{center}
			\caption{\label{tab1} Robust features of low-energy Aharonov-Bohm effect in the bent inhomogeneous superconducting wire}
	\begin{tabular}{ |c|c|c|c| } 
		\hline
		& AB period & Extrema positions, $\phi$ \\
		\hline
		inhomogeneous Andreev bound states & $\pi$ & $\pi n/2$ \\
		\hline
		bulk Andreev bound state & $2\pi$ & $\pi n$ \\
		\hline
		Majorana bound state & $2\pi$ & $\pi n/2$ \\
		\hline 
	\end{tabular}
\end{center}
\end{table}

Thus, the AB effect makes it possible to distinguish between all three types of excitations. Here it is important to emphasize once again that both characteristics can be the same for the different states due to fine-tuning of the parameters. For example, besides the already-noticed halving of the AB period in the topological phase at $A=0$ (see the expression \eqref{G_topo0}), the extrema of $G_{bulk}\left(\phi\right)$ may accidentally appear at $\phi=\pi \left(n+1/2\right)$ if $c_{1}b_{2}=c_{2}b_{1}$ (in addition to the parameter-independent extrema at $\phi=\pi n$) or the AB period for the bulk ABS becomes equal to $\pi$ if $b_{1,2}=0$.  

\subsection{\label{sec3.2} Effect of inhomogeneity}

At the end of this Section we consider the special cases concerning the coupling parameters $\tau_{i\sigma}$.
First, in the case of symmetric couplings, i.e. when $\tau_{1\left(2\right)\sigma}=\tau_{3\left(4\right)\sigma}$ and $\Gamma_{12}=\Gamma_{34}=\Gamma$, $\Delta\Gamma_{12}=\Delta\Gamma_{34}=\Delta\Gamma$, $\Delta\overline{\Gamma}_{12}=\Delta\overline{\Gamma}_{34}=\Delta\overline{\Gamma}$, $\Delta\overline{\Gamma}_{13}=\Delta\overline{\Gamma}_{24}=0$, the expressions \eqref{G_triv} and \eqref{G_topo} take a simpler form,
	\begin{equation} \label{G_triv_sym}
	G_{triv}^{sym}=4G_{0}\frac{\Delta\overline{\Gamma}^{2}\left(A^2+B^2+2AB\cos2\phi\right)}{\left(AB-\Delta\overline{\Gamma}^{2}\sin^{2}\phi\right)^{2}+\left(\Gamma^{2}+\Delta\overline{\Gamma}^{2}\right)\left(A+B\right)^{2}-2AB\Delta\Gamma^{2}},
	\end{equation}
	\begin{equation} \label{G_topo_sym}
	G_{topo}^{sym}=4G_{0}\frac{\Delta\overline{\Gamma}^{2}\left(A^2+B^2+2AB\cos\phi\right)}{\left(\Gamma^{2}+\Delta\overline{\Gamma}^{2}\right)\left(A+B\cos\phi\right)^{2}+B^{2}\left(A^{2}+\Delta\overline{\Gamma}^{2}\sin^{2}\phi\right)}.
	\end{equation}

Second, in the situation of the inhomogeneous device (Fig. \ref{1}b) the MC localization significantly depends on the smoothness degree of $V,~\Delta$-profiles \cite{penaranda-18,vuik-19}. In the setup under consideration as we will show below, the larger $\sigma_{i}$, the stronger one of the Majorana wave functions is localized in the inhomogeneous region. If, for example, $\sigma_{2}\gg\sigma_{1}$ (a 'steep-gentle' case described in Sec. \ref{sec4.2}) then in the trivial phase the probability density of the first MC, $M_{1}$, becomes small near the bottom edge of the device and $\tau_{1\sigma}\ll\tau_{2-4\sigma}$ (see Fig. \ref{4}c). Therefore, the trivial conductance equals (if $A\neq0$)
	\begin{equation} \label{G_triv_sg1}
	G_{triv1}^{s-g}\approx4G_{0}\frac{\Delta\overline{\Gamma}_{34}\left(A\Delta\overline{\Gamma}_{34}+2B\Delta\overline{\Gamma}_{12}\cos2\phi\right)}{A\left(B^2+\Gamma_{34}^{2}+\Delta\overline{\Gamma}_{34}^{2}\right)+2B\left(\Gamma_{12}\Gamma_{34}-\Delta\Gamma_{13}\Delta\Gamma_{24}-\Delta\overline{\Gamma}_{12}\Delta\overline{\Gamma}_{34}\sin^{2}\phi\right)}.
	\end{equation}
The other possible situation here is when the MC of the higher-energy ABS is confined in such a wide inhomogeneous area, e.g., $\tau_{3\sigma}\ll\tau_{1,2,4\sigma}$, resulting in
	\begin{equation} \label{G_triv_sg2}
	G_{triv2}^{s-g}\approx4G_{0}\frac{\Delta\overline{\Gamma}_{12}\left(B\Delta\overline{\Gamma}_{12}+2A\Delta\overline{\Gamma}_{34}\cos2\phi\right)}{B\left(A^2+\Gamma_{12}^{2}+\Delta\overline{\Gamma}_{12}^{2}\right)+2A\left(\Gamma_{12}\Gamma_{34}-\Delta\Gamma_{13}\Delta\Gamma_{24}-\Delta\overline{\Gamma}_{12}\Delta\overline{\Gamma}_{34}\sin^{2}\phi\right)}.
	\end{equation}
	
	Finally, when $\sigma_{1,2}\gg1$ (a 'gentle-gentle' case) one can expect the relation $\tau_{1,4\sigma}\ll\tau_{2,3\sigma}$ to be fulfilled and we find
	\begin{equation} \label{G_triv_gg}
	G_{triv}^{g-g}\approx4G_{0}\frac{A^2\Delta\overline{\Gamma}_{34}^{2}+B^2\Delta\overline{\Gamma}_{12}^{2}+2AB\Delta\overline{\Gamma}_{12}\Delta\overline{\Gamma}_{34}\cos2\phi}{\Theta_{triv}}.
	\end{equation}
	It follows from \eqref{G_triv_sg2} and \eqref{G_triv_gg} that in the case of zero-energy ABS ($A=0$) the last two expressions become equivalent and do not depend on the magnetic flux. In opposite, the AB oscillations persist when $\tau_{1\sigma}\ll\tau_{2-4\sigma}$ if $\Delta\overline{\Gamma}_{12}\Delta\overline{\Gamma}_{34}\neq0$.
\end{widetext} 

\begin{figure*}[!htb]
	\begin{center}
		\includegraphics[width=0.465\textwidth]{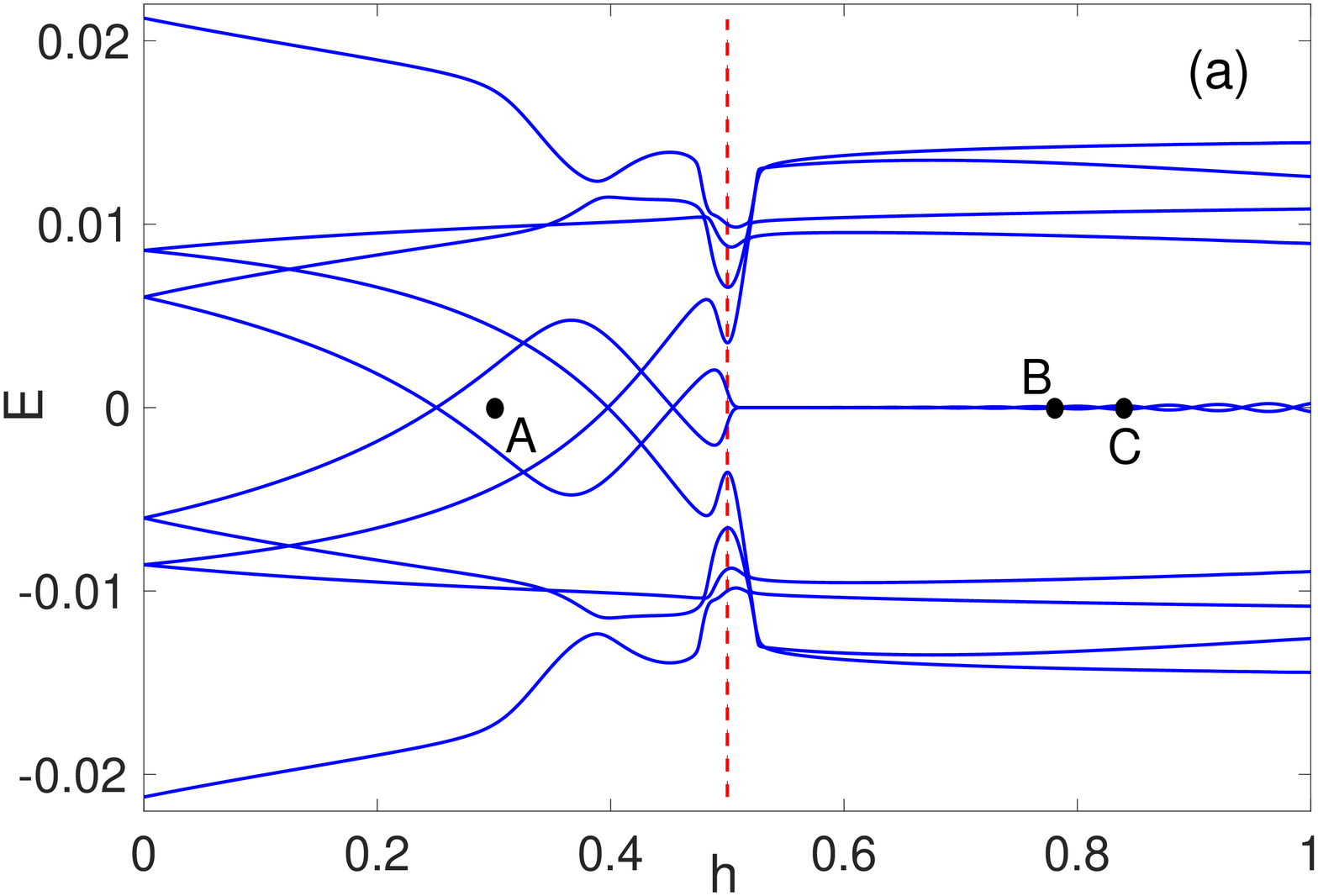}
		\includegraphics[width=0.465\textwidth]{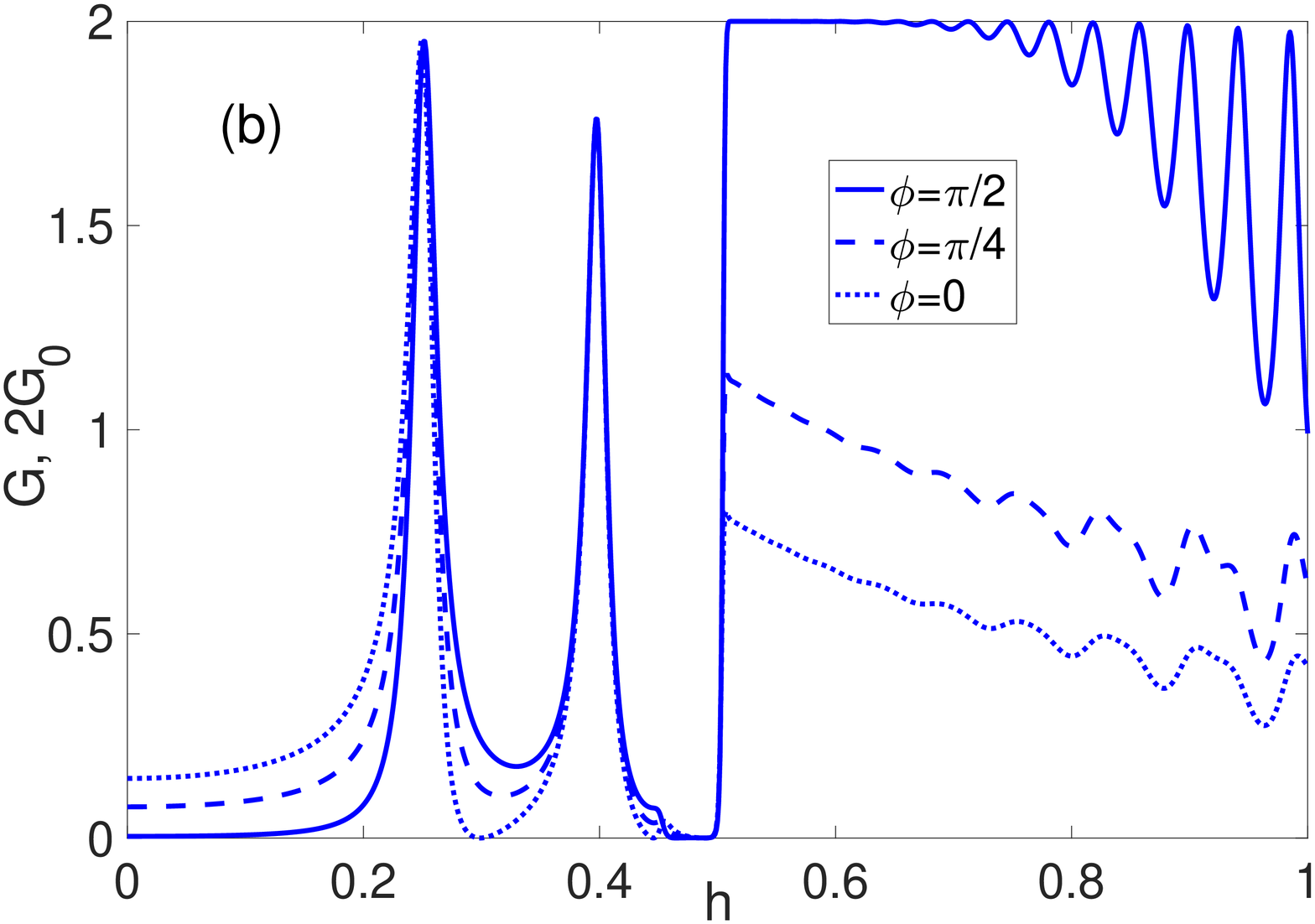}
		\includegraphics[width=0.465\textwidth]{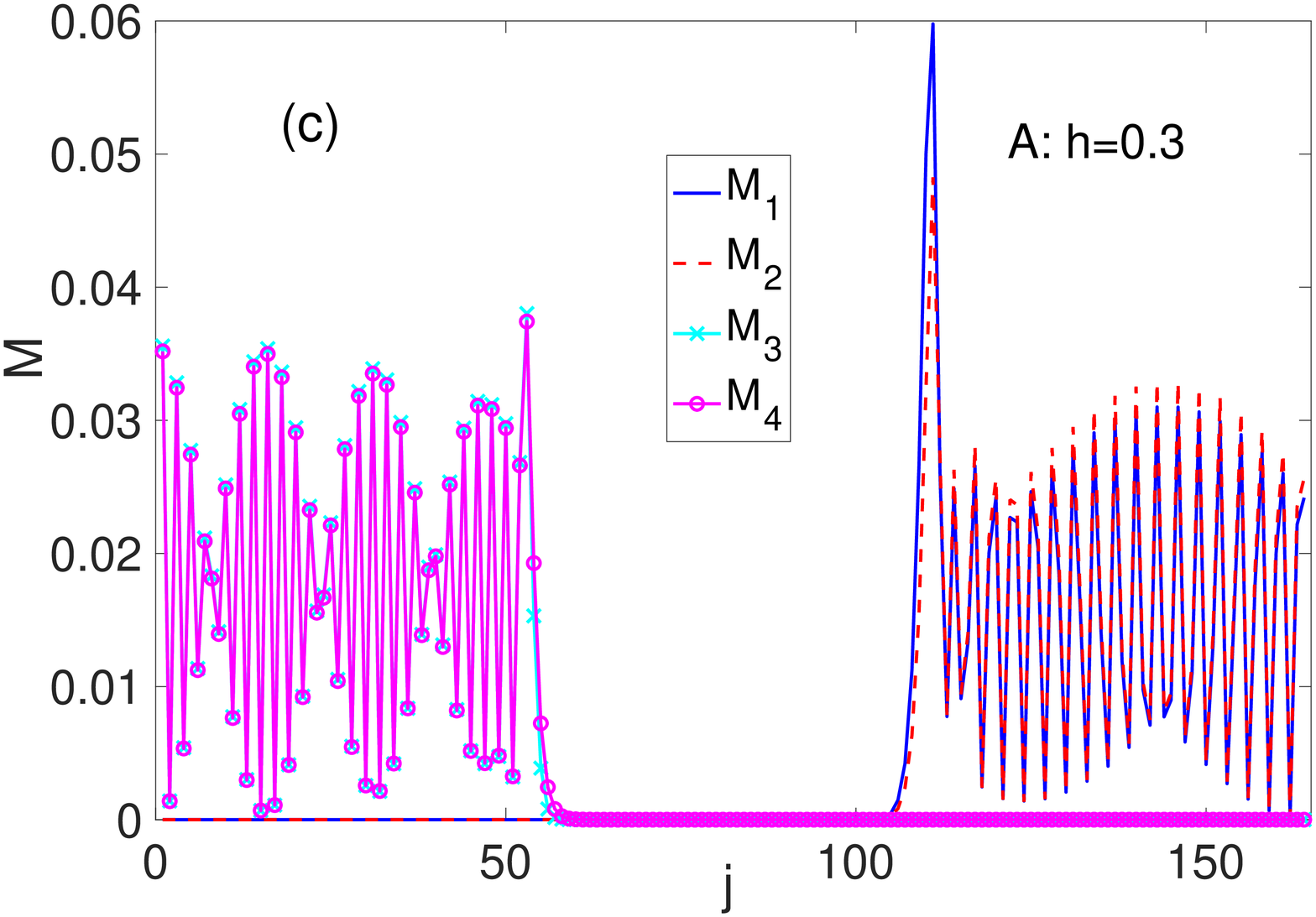}
		\includegraphics[width=0.465\textwidth]{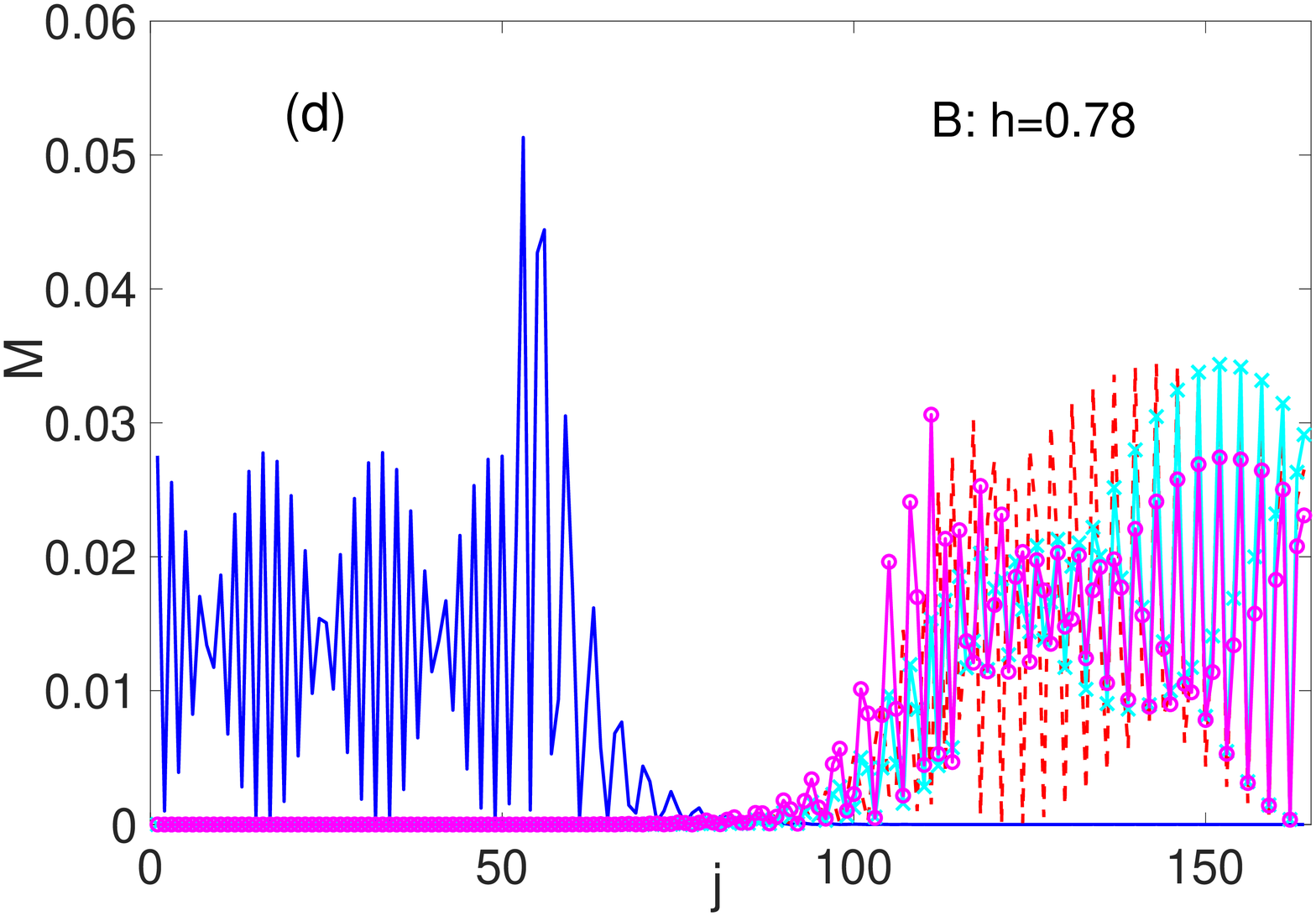}
		\caption{\label{2} Low-energy part of spectrum (a) and linear-response conductance (b) of the interference device as functions of the Zeeman energy for smooth steep change of electrostatic and superconducting pairing potentials at both N1/S and S/N2 interfaces. Spatial distributions of the probability densities of Majorana components for the first (plotted w/t markers) and second (plotted with markers) lowest-energy excitations in the trivial (c) and nontrivial (d) phases. Parameters: $\sigma_{1}=2, \sigma_{2}=3$.}
	\end{center}
\end{figure*}
 
\section{\label{sec4}Results and discussion}

Below we focus on the spectral and transport properties of the inhomogeneous interference device obtained numerically. The linear-response conductance is calculated using the nonequilibrium Green's function method. The details of this technique in the tight-binding approximation are presented in \cite{valkov-22}. Unless otherwise specified, the following parameters are used in the calculations: $N_S=60, N_1=N_2=52, \varepsilon_1=-0.1, \varepsilon_S=1.4, \varepsilon_2=-0.5, t_{1,2\sigma}=1, \Delta_{0}=0.3, \alpha_{1}=-\alpha_{2}=0.3, \alpha_{S}=0.2, \mu=0, \Gamma_{1,2}=0.04$. The energy quantities are measured in units of $t=\hbar^2/ma^2\approx 2~meV$ where $a=50$ nm and $m=0.015m_{0}$ in order to model the experimentally investigated InAs, InSb wires of micrometer size. Zeeman-energy dependencies of conductance discussed in Subsecs. \ref{sec4.1}, \ref{sec4.2} as well as conductance maps in Subsec. \ref{sec4.3} are plotted in the zero temperature limit, $kT\approx0$. 

\subsection{\label{sec4.1} 'Steep-steep' case}

We start with the situation where the $V\left(j\right)$ and $\Delta\left(j\right)$ functions are smooth at both interfaces and the corresponding changes are steep, $\sigma_{1}=2$, $\sigma_{2}=3$ (see dashed curves in Fig. \ref{1}b). A low-energy part of the device spectrum as a function of the Zeeman energy is depicted in Fig. \ref{2}a. The local nature of the in-plane magnetic field leads to the presence of states whose energies weakly depend on $h$. A vertical dashed line, $h=h_{c1}$, indicates a lower border of the topologically nontrivial phase of the S-segment, $h_{c1,2}=\sqrt{\Delta_{0}^2+\left(\varepsilon_{S}-\mu\mp t\right)^2}$. Note that even if it is in the trivial phase, the zero-energy inhomogeneous ABSs occur provided by the two pairs of the Bogolyubov excitations (instead of the one pair in the case of single smooth inhomogeneity \cite{moore-18,penaranda-18}) and inhabit only a certain part of the device.

A typical spatial distribution of their MC probability densities, $M_{1-4}$, is depicted in Fig. \ref{2}c for the Zeeman energy marked with point 'A' in Fig. \ref{2}a. The Majorana wave functions of the first trivial ABS are localized inside the N2-section and in the SC part close to the S/N2 interface where, in particular, both $M_{1}$ and $M_{2}$ (displayed by solid and dashed curves, respectively) have the pronounced maxima. 

As was already emphasized in the analysis of the expression \eqref{G_triv}, in the trivial phase, for a complete and correct description of interference effects, it is necessary to take into account the presence of the second ABS, which can also have a low energy and is localized in the opposite arm of the device. The probability densities related to the second trivial ABS, $M_{3}$ and $M_{4}$, are drawn in Fig. \ref{2}c by the curves with markers 'x' and 'o', respectively. The Majorana wave functions dwell in the N1-section and in the SC part near the N1/S interface. Thus, the trivial ABSs (with both zero and nonzero energy) are characterized by a strong overlap of the MCs wave functions. These MCs interact with the contact only through one of the device arms. Note the observed features of spectrum and Majorana wave functions resemble those found in the simplier system with a single quantum dot attached to the SC wire \cite{liu-17b,penaranda-18}. This is a typical structure studied in modern tunnel spectroscopy experiments with the hybrid nanowires pursuing the MBS detection \cite{yu-21}.
 
In Fig. \ref{2}b the $h$-dependence of conductance is displayed. The trivial zero-energy ABSs lead to resonant peaks in $G\left(h\right)$ which can be close to the quantized value $4G_{0}$. As follows from the numerator of \eqref{G_triv}, in the most common situation, the antiresonances arise as a result of destructive interference involving the transport channels associated with both ABSs (although the Fano effect mediated by the two MCs localized in the same arm is also possible). Consequently, such antiresonances are $\phi$-dependent. This behavior is shown in Fig. \ref{2}b where the Fano antiresonance appears at $h=0.3$ if $\phi=0$ (dotted curve) and is absent if $\phi=\pi/4,~\pi/2$ (dashed and solid curves, respectively). In turn, the AB phase does not affect the trivial peaks, which agrees with the formula \eqref{G_triv0} for $B \gg A,~\Gamma_{i\sigma,j\sigma'}$ (for the chosen parameters $B\sim10^{-3}$, $\Gamma_{i\sigma,j\sigma'}\sim 10^{-4}$).  

After the topological phase transition ($h>h_{c1}$) the pair of eigenstates converges to zero energy. The corresponding spatial distributions of $M_{1}$ and $M_{2}$ plotted in Fig. \ref{2}d become separated and are localized in the opposite parts of the structure (compare the solid as well as dashed curves in Figs. \ref{2}c,d). One can see that the MCs leak out from the SC wire opposite ends into the adjacent normal sections \cite{klinovaja-12,vernek-14}. Hence, at $h>h_{c1}$ the MCs are coupled to the contact through the different arms which makes it possible to implement the AB effect even without the additional consideration of tunneling into the second state. This provides the main difference between the trivial ABS and topological MBS. Note that since the SC-wire Hamiltonian belongs to symmetry class D \cite{altland-97,kitaev-09} with the only possible MBS, the $M_{3}$ and $M_{4}$ are still present in the same part of the device as seen from Fig. \ref{2}d.

\begin{figure*}[!htb]
	\begin{center}
		\includegraphics[width=0.475\textwidth]{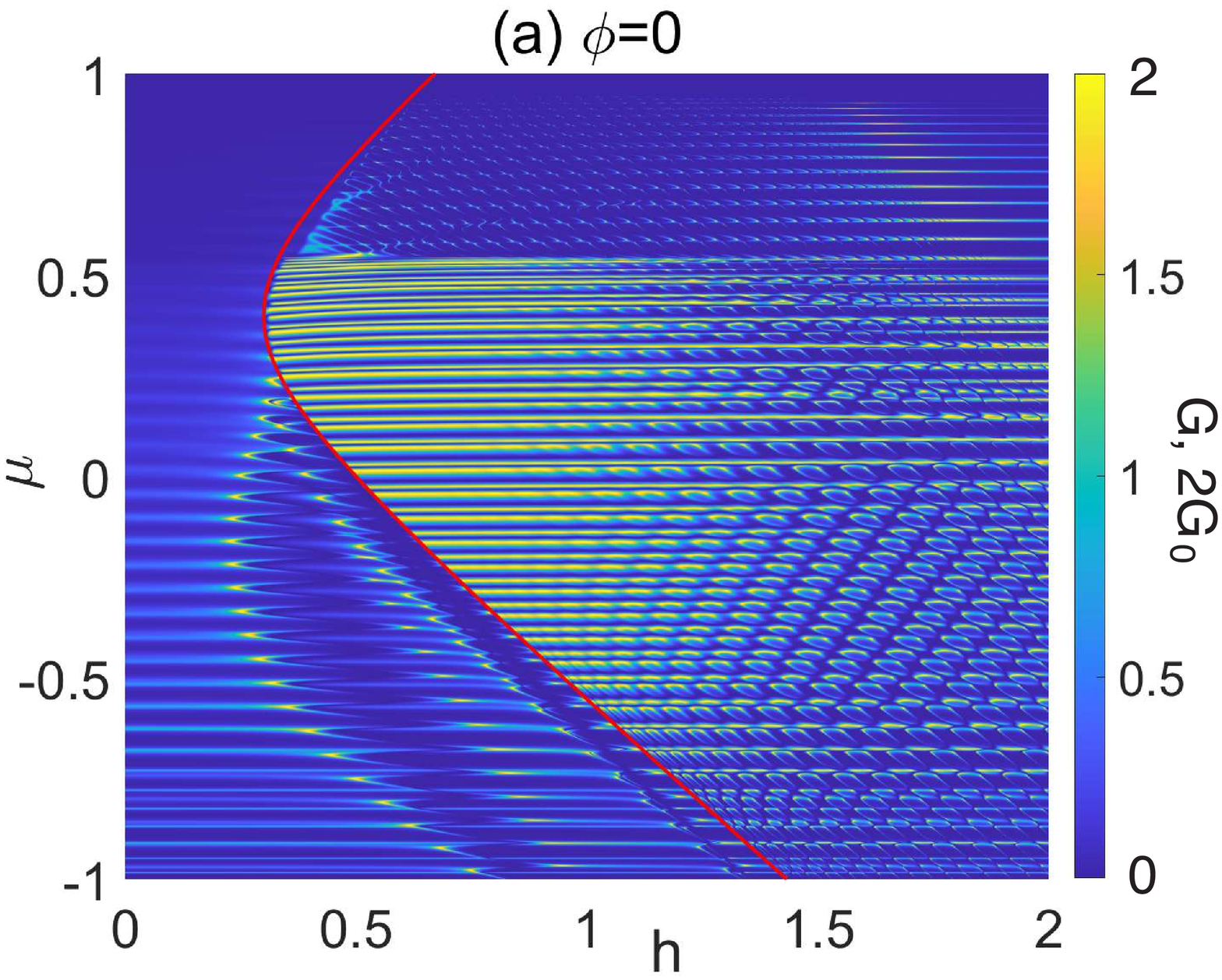}
		\includegraphics[width=0.475\textwidth]{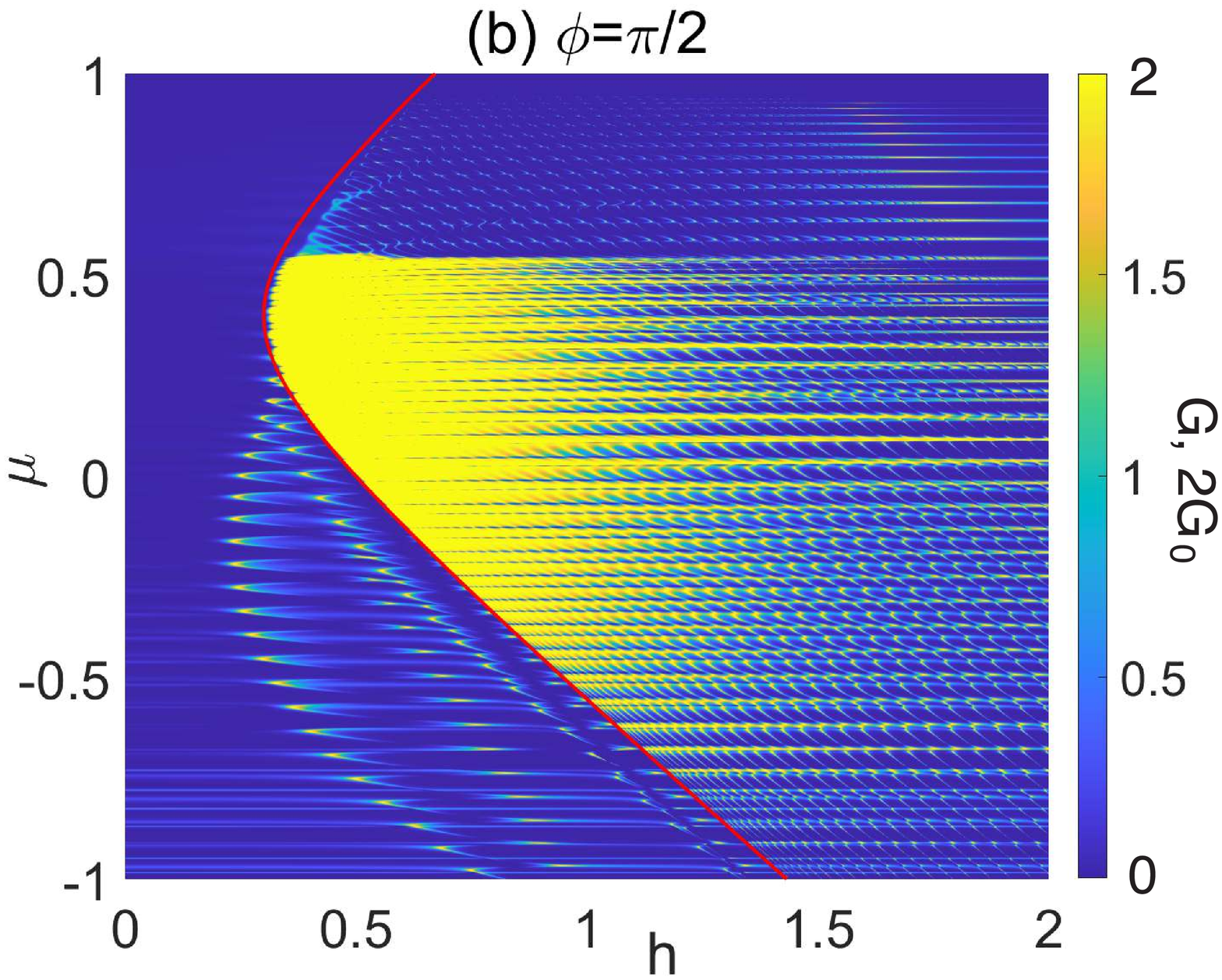}
		\includegraphics[width=0.475\textwidth]{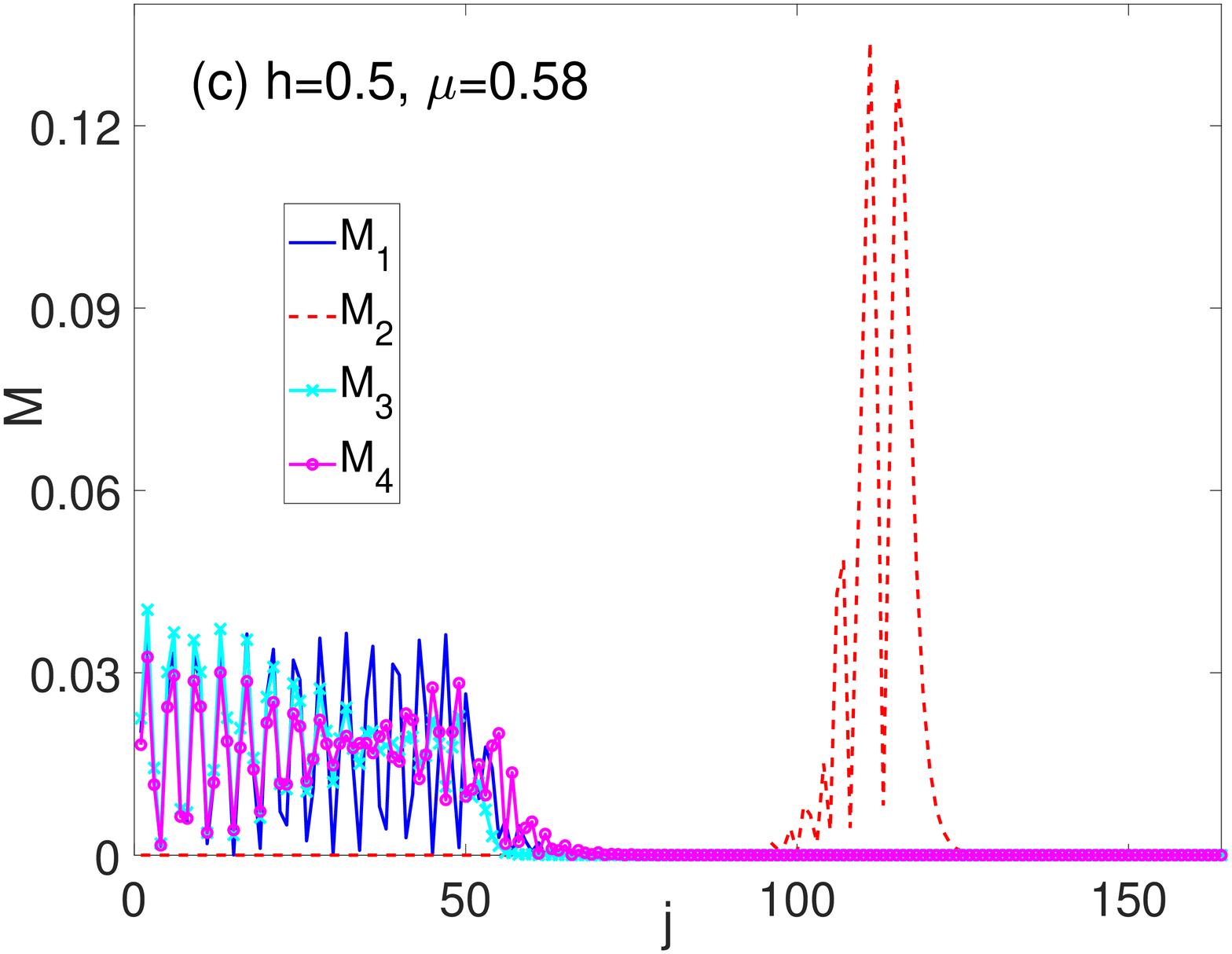}
		\includegraphics[width=0.475\textwidth]{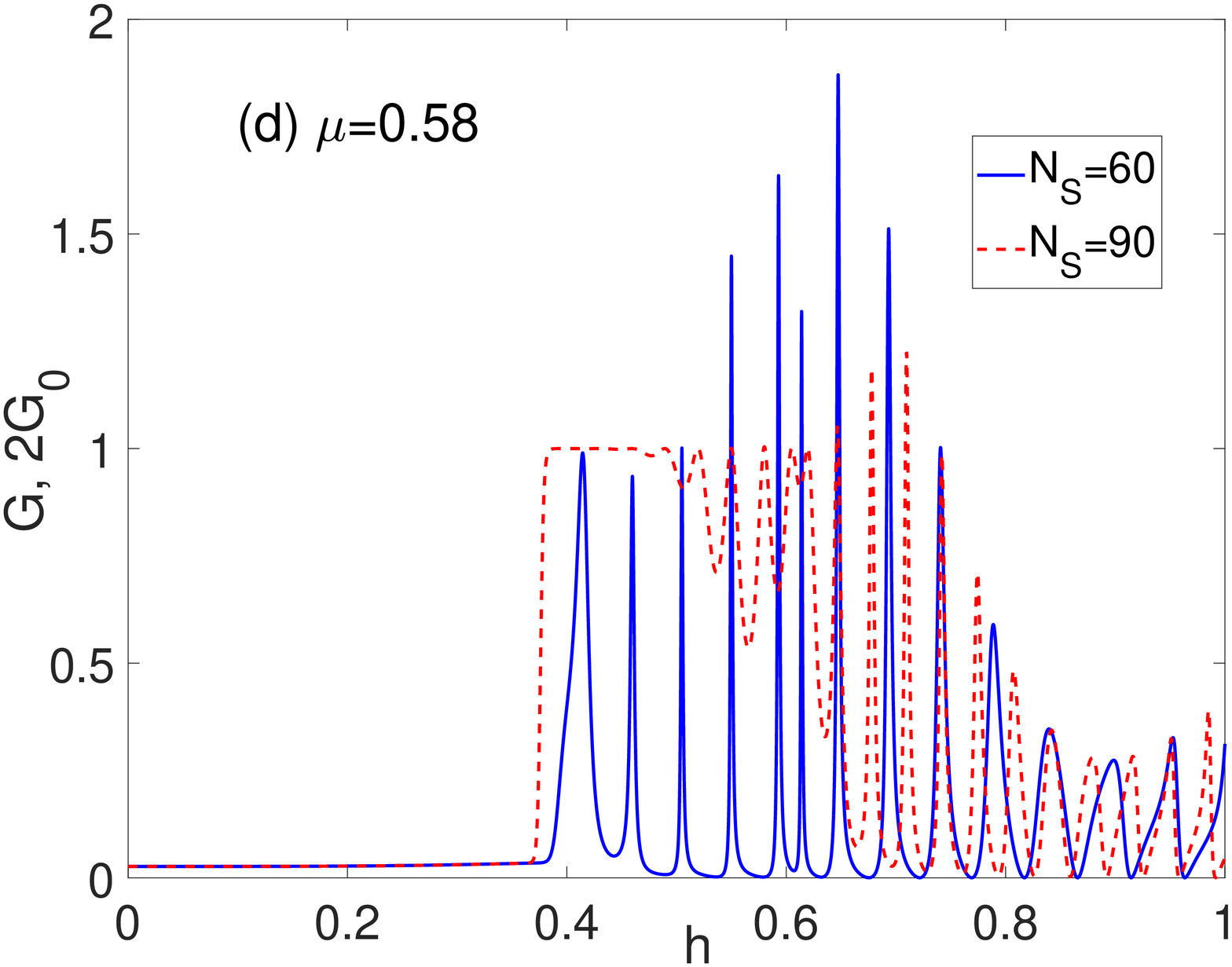}
		\caption{\label{5} Conductance maps $G\left(h,~\mu\right)$ at (a) $\phi=0$ and (b) $\phi=\pi/2$ for the 'steep-steep' type of the $V\left(j\right),~\Delta\left(j\right)$ profiles. The red solid curves indicate the boundary between the trivial (left) and nontrivial (right) phases of the S-segment in the $\left(h,~\mu\right)$ parametric space. (c) Spatial distributions of the probability densities of Majorana components for the first (plotted w/t markers) and second (plotted with markers) lowest-energy excitations at $h=0.5$, $\mu=0.58$. (d) Zeeman-energy dependence of conductance at $\mu=0.58$.}
	\end{center}
\end{figure*}

According to \eqref{G_topo0}, immediately after the topological phase transition, the conductance can be amplified by the magnetic flux and reaches a $4G_{0}$-height plateau at $\phi=\pi/2$ (see Fig. \ref{2}b) \cite{liu-11,aksenov-22}. Then, as the Zeeman energy grows the conductance starts to oscillate with the increasing amplitude due to the overlap of MC wave functions. Thus, the zero-energy trivial ABSs and nontrivial MBSs can be distinguished by the different behavior of the function $G\left(\phi\right)$. Interestingly, since the MC leakage into the arms is sustainable to the change of the model parameters, one can efficiently control the interference transport, e.g., varying the on-site energy of the normal section (see Figs. \ref{5}a,b and Sec. \ref{sec4.3}).

\begin{figure}[!tb]
	\begin{center}
		\includegraphics[width=0.5\textwidth]{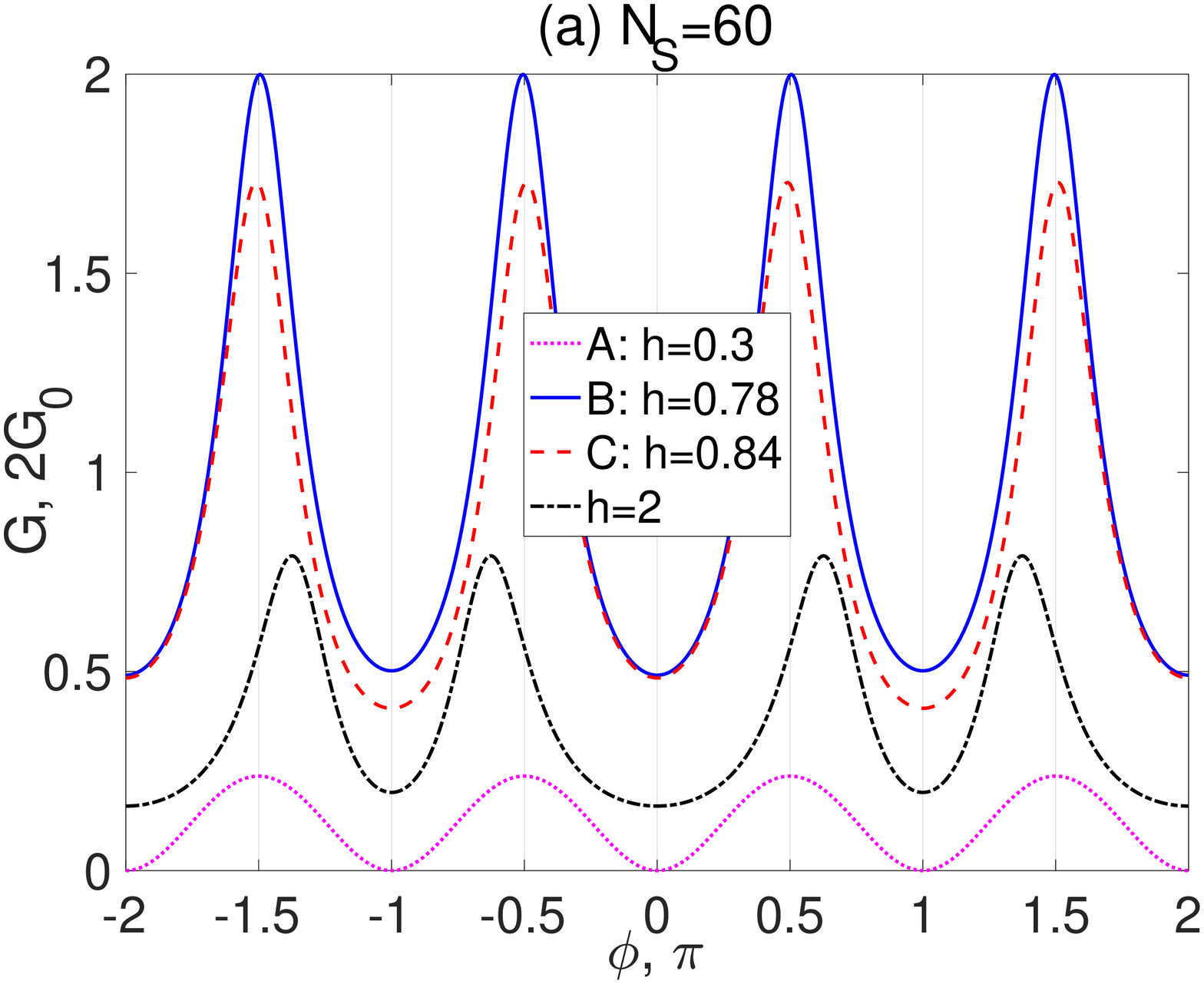}
		\includegraphics[width=0.5\textwidth]{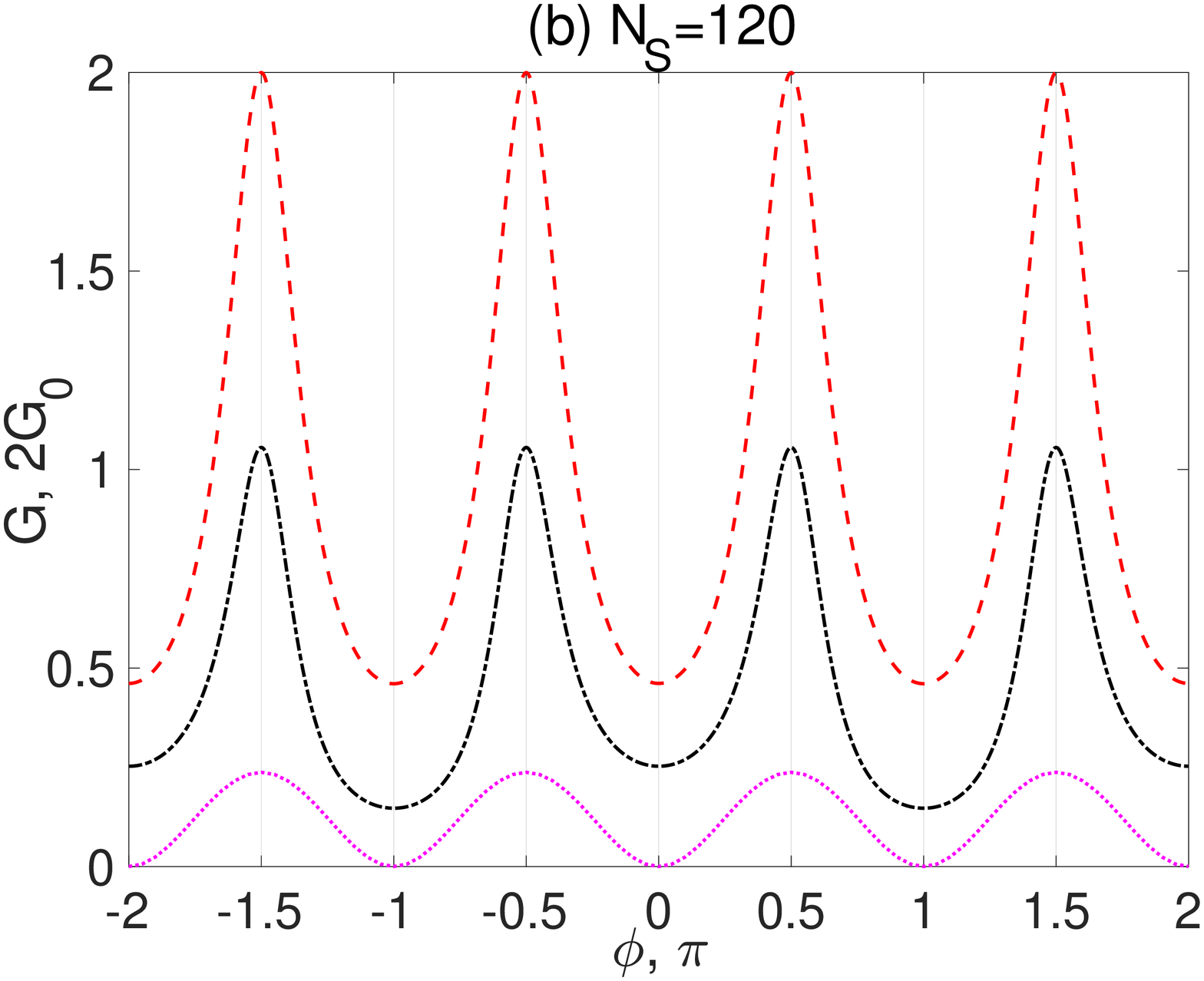}
		\caption{\label{AB_steep-steep} The Aharonov-Bohm oscillations of conductance if the smooth inhomogeneity has the steep spatial dependence at both normal metal/superconductor interfaces. The Zeeman energy of the point 'A' ('B' and 'C'), indicated in Fig. \ref{2}a, corresponds to the trivial (nontrivial) phase of the S-section. (a) $N_{S}=60$, (b) $N_{S}=120$. Parameters: $\sigma_{1}=2, \sigma_{2}=3$.}
	\end{center}
\end{figure}

The AB oscillation period provides another distinction between the ABS and MBS. As was noticed in Sec. \ref{sec3} the $G\left(\phi\right)$ is always $\pi$-periodic in the trivial phase (see dotted curve in Fig. \ref{AB_steep-steep}a) whereas the periodicity after the topological phase transition essentially depends on the $A$ value. In other words, it correlates with the $A$ oscillations as one can see comparing the solid ($A\approx0$) and dashed ($A\neq0$) curves in Fig. \ref{AB_steep-steep}a. Since in practice the situation of the MBS with nonzero energy is more likely to arise, the different periodicity of the AB oscillations is a suitable feature for distinguishing between the trivial and nontrivial subgap states. 

However, as we also mentioned above, at the high magnetic, when $M_{1}$ ($M_{2}$) penetrates sufficiently into the bottom (top), the bulk ABS emerges (formally in the nontrivial phase) and the AB period, in general, is $2\pi$ even though $A\approx0$. The latter takes place since each Majorana mode couples with the normal contact via both arms \cite{aksenov-22}. Such a situation at $h=2$, near the upper border ($h=h_{c2}$) of the S-segment topological phase, is depicted by dash-dotted curve in Fig. \ref{AB_steep-steep}a. According to the results of Table \ref{tab1}, in order to successfully separate these three situations, it is also necessary to pay attention to the extrema positions in $G\left(\phi\right)$. In particular, the AB interference based on the MBS causes the parameter-independent maxima and minima at $\phi=\pi n/2$ as displayed by solid and dashed curves. The trivial inhomogeneous ABSs possess the same feature shown by the dotted curve. In opposite, such robust extrema for the bulk ABS appear only if $\phi=\pi n$, while the positions of the extrema between them can be changed by varying the system parameters.

\begin{figure*}[!tb]
	\begin{center}
		\includegraphics[width=1\textwidth]{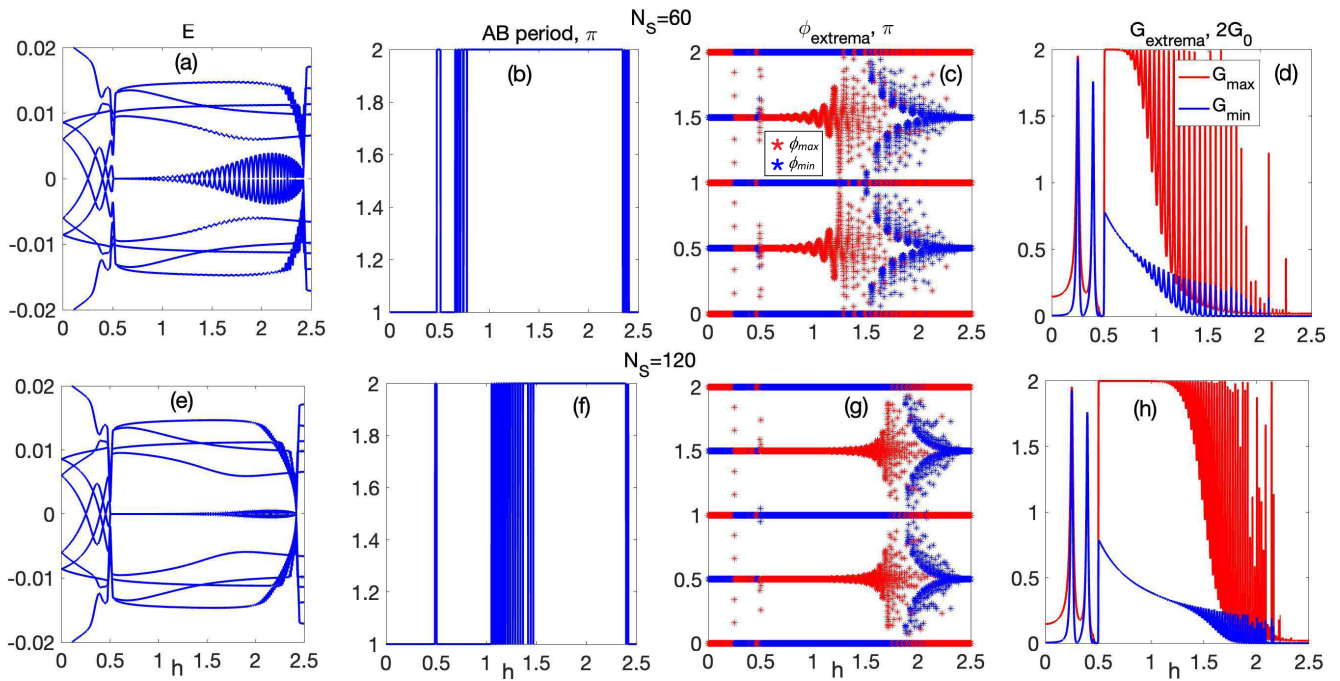}
		\caption{\label{AB_set1} Properties of the low-energy Aharonov-Bohm effect in the bent inhomogeneous superconducting wire as function of the Zeeman energy for $N_{S}=60$ (a-d) and $N_{S}=120$ (e-h). (a,e) Excitation spectrum; (b,f) Aharonov-Bohm period; (c,g) Aharonov-Bohm phase of the conductance maxima (red stars) and minima (blue stars); (d,h) values of conductance maxima (red) and minima (blue). Parameters: $\sigma_{1}=2, \sigma_{2}=3$.}
	\end{center}
\end{figure*}

In Fig. \ref{AB_steep-steep}b the AB oscillations are plotted when the SC-wire length is doubled ($N_S=120$). It can be seen that the AB effect induced by the inhomogeneous ABSs is not modified in this situation since these excitations are caused by the features of the N/S interfaces and are predominantly localized in the N-sections. In turn, since the bulk ABS turns into the MBS the robust extrema appear at $\phi=\pi\left(n+1/2\right)$ in addition to the extrema at $\phi=\pi n$, while the $2\pi$ period does not change. Interestingly, the oscillations due to the MBS with initially nonzero energy are transformed from $2\pi$- to $\pi$-periodic and the conductance maxima acquire quantized values (compare dashed curves in Figs. \ref{AB_steep-steep}a,b). The effect is explained by the decrease in the MBS energy by two orders of magnitude, which actually leads to the zero-energy MBS (this case is also shown by solid curve in Fig. \ref{AB_steep-steep}a).

The gradual evolution of the AB effect properties as the Zeeman energy is swept is presented in Fig. \ref{AB_set1}. The top (bottom) row corresponds to the SC wire length $N_{S}=60$ ($N_{S}=120$). The data are in full agreement with the analytical results collected in Table \ref{tab1}. At the low magnetic fields $h<0.5$, when the SC wire is in trivial phase, the AB period is $\pi$ (Fig. \ref{AB_set1}b) and the robust conductance minima and maxima appear at $\phi=\pi n/2$ (see blue and red stars in Fig. \ref{AB_set1}c, respectively). Immediately after the topological phase transition, in a small range of $h$ the period does not change since $A\approx0$. Here, the maximum conductance in the AB effect is quantized on the $4G_{0}$ plateau, while the minimum one monotonically decreases (see red and blue solid curves in Fig. \ref{AB_set1}d). Then, oscillations of the AB period are observed, which are unique for the MBS case, as predicted by the formulas \ref{G_topo} and \ref{G_topo0}. Simultaneously, $G_{max}\left(h\right)$ and $G_{min}\left(h\right)$ also start to oscillate. It can be stated with good accuracy that in both these Zeeman-energy regions ($0.5<h<1$) the conductance extrema arise at $\phi=\pi n/2$. As $h$ increases further, the period stabilizes at $2\pi$. However, at $h>1$ the oscillations of $\phi_{max}$ around $\pi/2$ and $3\pi/2$ grow indicating the realization of the bulk ABS whose conductance extrema at $\phi=\pi\left(n+1/2\right)$ are parameter-dependent. The described behavior does not fundamentally change for the longer S-section. The AB-period oscillations in the nontrivial phase shift to the higher values of $h$ (Fig. \ref{AB_set1}f). Hence, the bulk ABS occurs now at $1.5<h<2.4$ where the AB phases of the robust conductance extrema are only $\phi=\pi n$. To conclude this Section, it is worth to note that near the critical energies $h_{c1}\approx0.5$, $h_{c2}\approx2.4$ the pronounced changes in the AB period are observed. Here the linear-response transport is determined by at least two bulk ABSs. The analysis of the AB effect in the immediate vicinity of the topological phase transitions is beyond the scope of this study.

\subsection{\label{sec4.2} Effect of profiles smoothness}

\begin{figure*}[!htb]
	\begin{center}
		\includegraphics[width=0.475\textwidth]{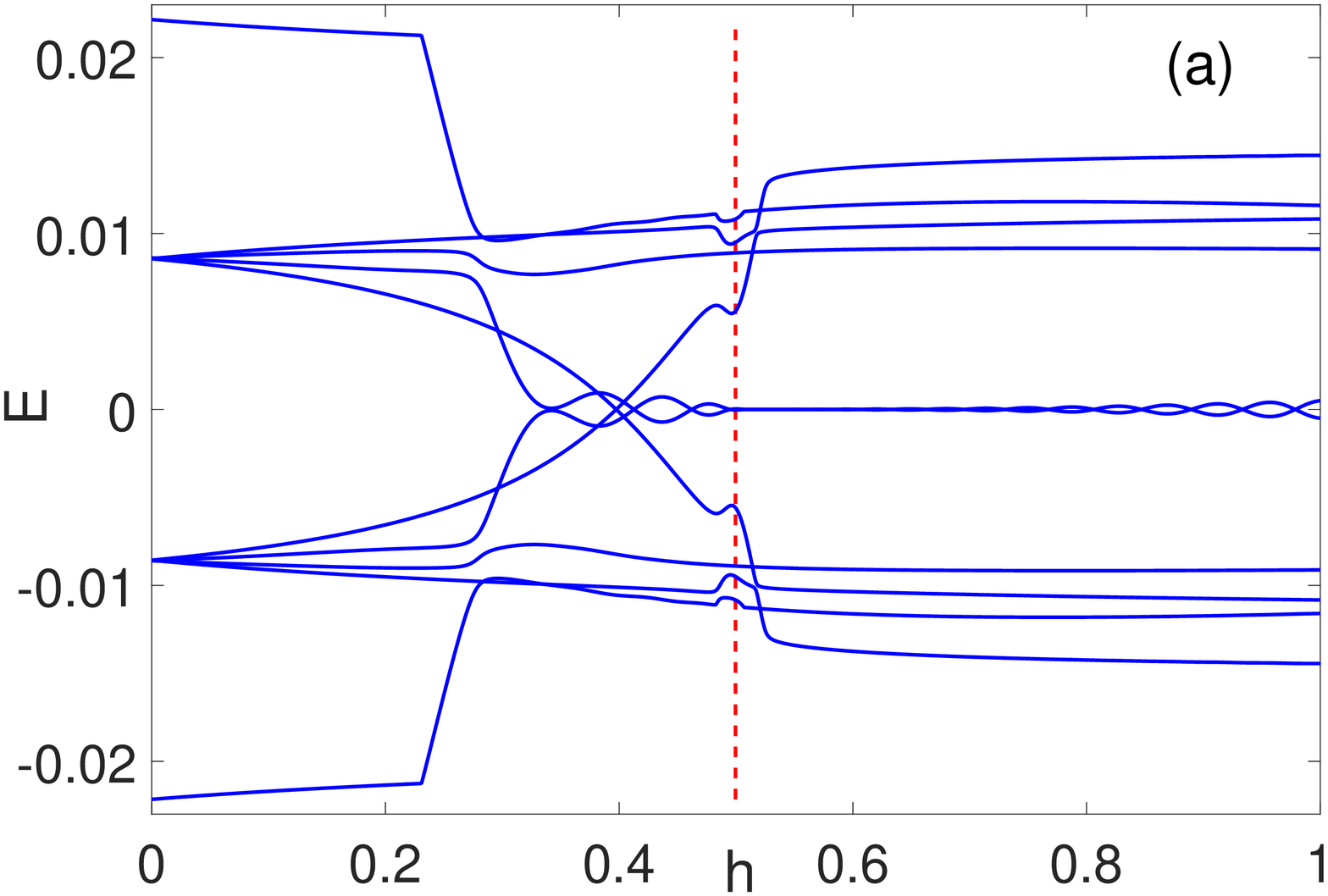}
		\includegraphics[width=0.475\textwidth]{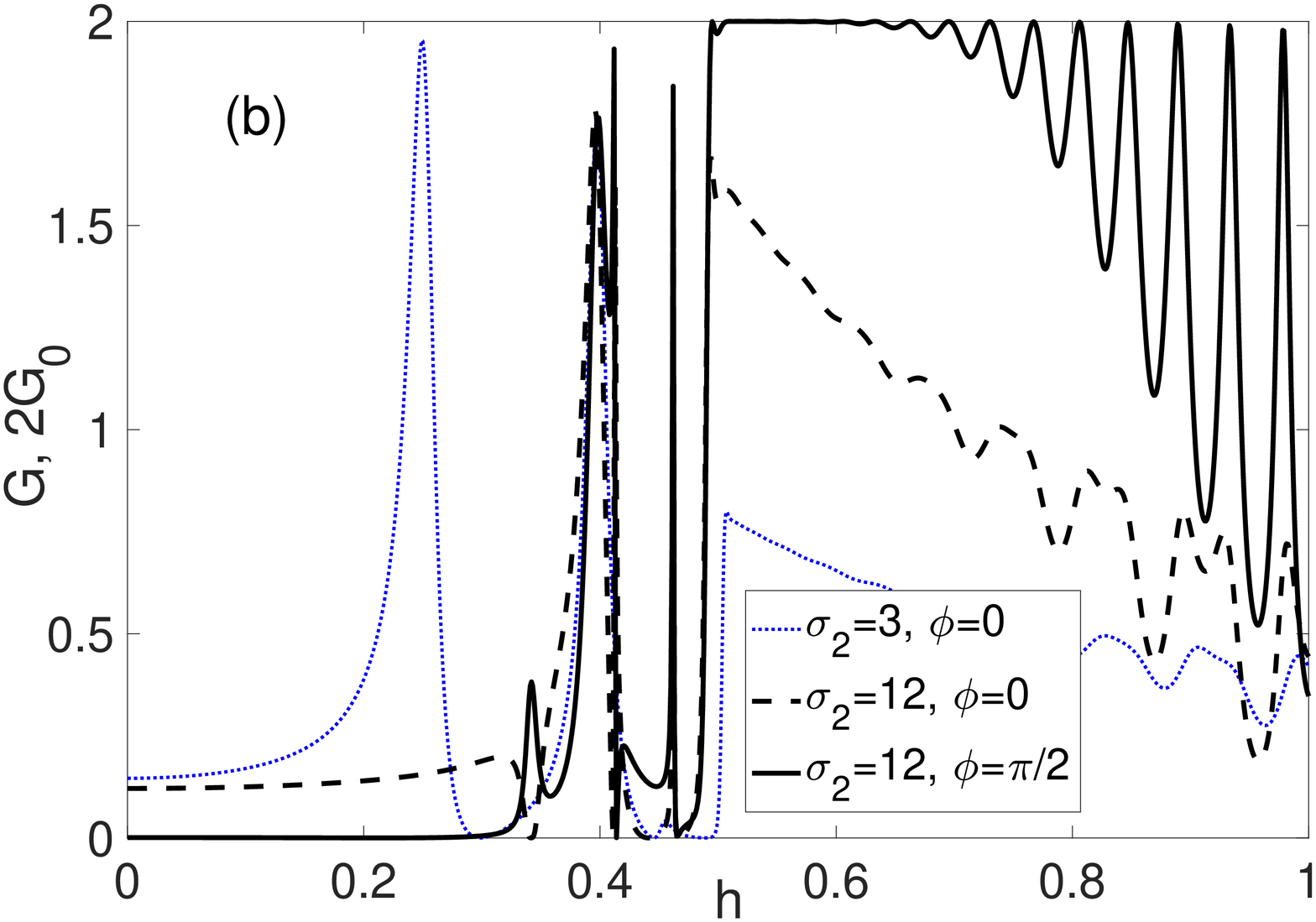}
		\includegraphics[width=0.475\textwidth]{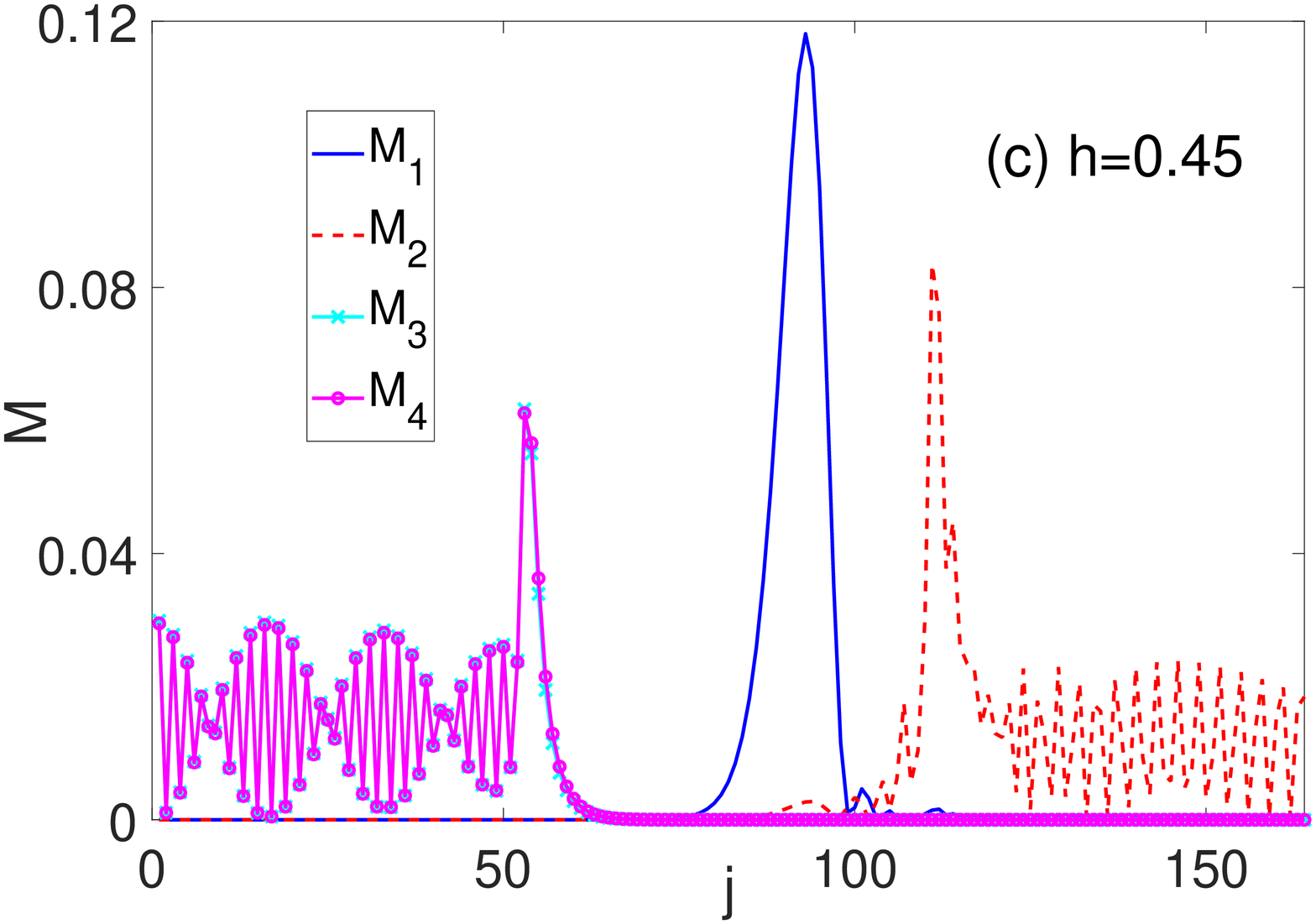}
		\includegraphics[width=0.475\textwidth]{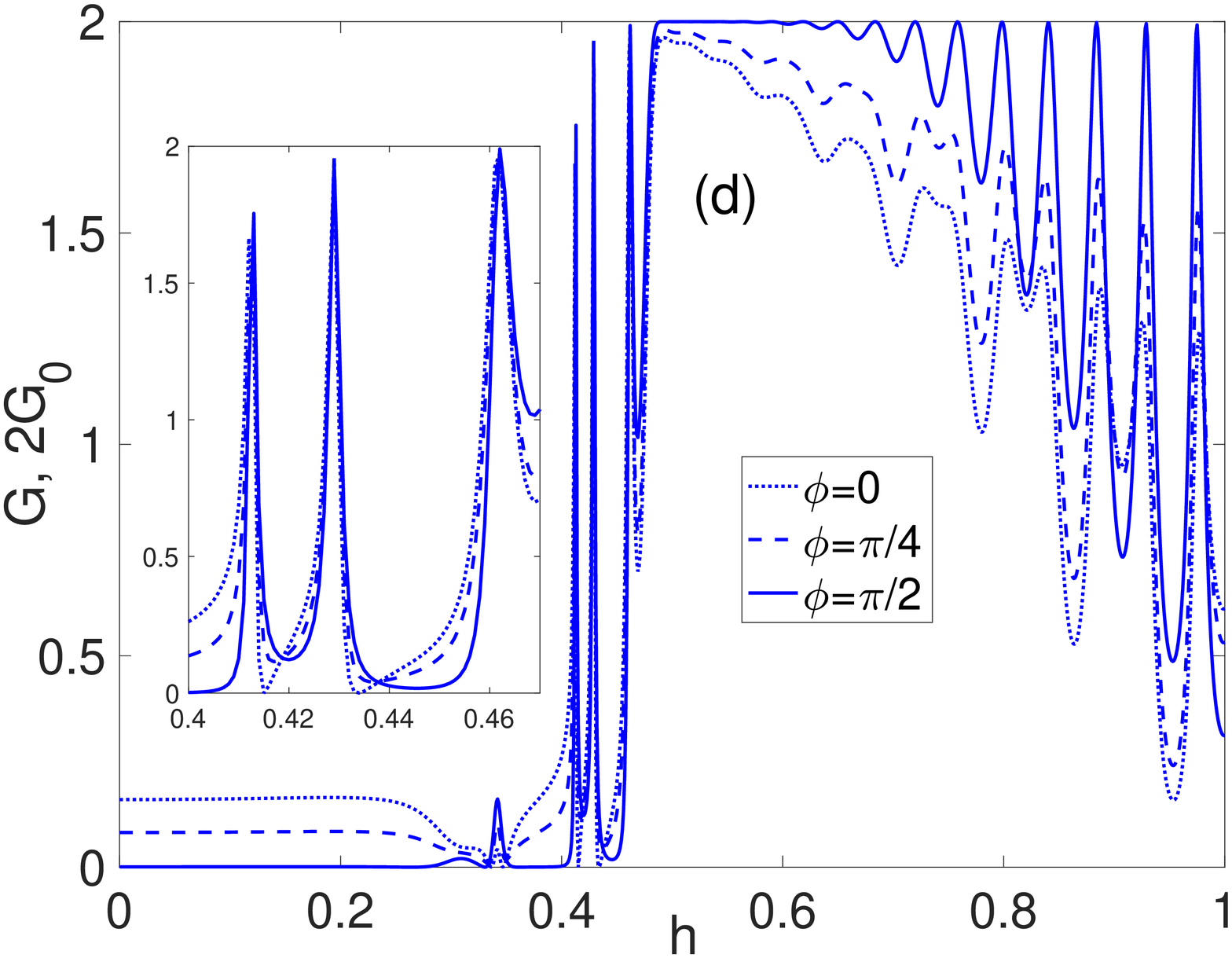}
		\caption{\label{4} Low-energy part of spectrum (a) and conductance (b) of the interference device as functions of the Zeeman energy for the steep and gentle change of electrostatic and superconducting pairing potentials at the N1/S ($\sigma_{1}=2$) and S/N2 ($\sigma_{2}=12$) interfaces, respectively.  (c) Spatial distributions of the probability densities of Majorana components for the first (plotted w/t markers) and second (plotted with markers) lowest-energy excitations in the trivial phase ($h=0.45$). (d) Zeeman-energy dependence of conductance in the 'gentle-gentle' case ($\sigma_{1}=8$, $\sigma_{2}=12$). Inset: the conductance peaks induced by quasi-MBSs in the higher $h$ resolution.
		}
	\end{center}
\end{figure*}

Next, let us consider the 'steep-gentle' case when the change of $V\left(j\right)$ and $\Delta\left(j\right)$ at the S/N2 interface is gentle ($\sigma_2=12$) while the smooth inhomogeneity at the N1/S boundary remains steep ($\sigma_1=2$). One can see from Fig. \ref{4}a that the energy of the ABS localized in the bottom part at $h<h_{c1}$ demonstrates decreasing oscillations near zero as the Zeeman energy increases. Meanwhile, the dependence on $h$ of the energy of the upper-arm ABS remains practically the same as in Fig. \ref{2}a.

The spatial distribution of the MC probability densities for this state is also similar to the 'steep-steep' case. In particular, the $M_{3}$ and $M_{4}$ strongly overlap, have the maxima at the N1/S boundary and are suppressed in the bulk of the S-segment and the N2-region (see marked curves in Fig. \ref{4}c). The ABS with the stabilized-near-zero energy includes the Majorana components which are significantly separated. The $M_{2}$ exhibits a peak in the normal part of the inhomogeneous area and oscillating behavior throughout the whole N2-region. On the contrary, the $M_{1}$ does not oscillate. It is localized in the SC part of the gentle inhomogeneity and overlaps with the second MC only slightly giving rise to the quasi-MBS. The separation of $M_{1}$ and $M_{2}$ is enhanced with the increase of $h$ that can be treated as a precursor of the topological phase transition since the first MC moves to the top arm.

In Fig. \ref{4}b the conductance for the 'steep-gentle' case (dashed and solid curves) is compared with the one for the 'steep-steep' case (dotted curve). In the trivial phase the peak at $h\approx0.4$ related to the zero-energy ABS which is induced by the steep inhomogeneity remains unchanged. Indeed, according to the formula \eqref{G_triv_sg2}, the resonance width is determined by the coupling parameters $\tau_{1,2\sigma}$ corresponding to the MCs of the upper arm. Additionally, this conductance maximum is not affected by the AB phase (compare dashed and solid curves). 

The resonances and antiresonances corresponding to the zero-energy quasi-MBSs are much narrower. It is in agreement with the expressions \eqref{G_triv0} and \eqref{G_triv_sg1} which dictate that the width of conductance features is proportional to the product of all four coupling coefficients $\tau_{i\sigma}$. Note that the magnetic-flux influence on these peculiarities is ambiguous since $B\sim\Gamma_{i\sigma,j\sigma'}$ at $h\approx0.4$ and the product $\Delta\overline{\Gamma}_{12}\Delta\overline{\Gamma}_{34}$ is $h$-dependent (see formula \eqref{G_triv0}).
After the topological phase transition the above mentioned enhancement of conductance by the magnetic flux persists.

In the 'gentle-gentle' case ($\sigma_{1,2}\gg1$), we obtain two Majorana components, predominantly localized in the different inhomogeneous regions at the N1/S and S/N2 interfaces, respectively, leading to $\tau_{1,4\sigma}\ll\tau_{2,3\sigma}$. As a result, all the resonant features in $G\left(h<h_c\right)$ displayed in Fig. \ref{4}d become narrow. Moreover, according to the expression \eqref{G_triv_gg} the influence of the AB phase on the peaks is negligible (see inset of Fig. \ref{4}d). In contrast, if the S-section is in the nontrivial phase the conductance maximum can still be controlled by the magnetic flux. 

It is worth emphasizing that in all three cases of the different profiles ('steep-steep', 'steep-gentle', and 'gentle-gentle') the numerically obtained period of the AB oscillations in the trivial phase is equal to $\pi$, while at $h>h_{c1}$ it is $2\pi$ (if the MBS energy is not exactly zero), which confirms the analytical results \eqref{G_triv} and \eqref{G_topo}. Such stability of the results allows us to consider the change in the periodicity of the AB oscillations as a sign indicating the topological phase transition in the inhomogeneous Majorana wires.

\subsection{\label{sec4.3} Conductance maps}

In Fig. \ref{5}a the conductance is plotted as a function of $h$ and $\mu$ at $\phi=0$ for the 'steep-steep' profiles of $V\left(j\right),~\Delta\left(j\right)$ . There are two areas separated by red solid curves, $$\mu=\pm\sqrt{h^2-\Delta_{0}^{2}}+\varepsilon_{S}-t,$$ 
where the $G\left(h,~\mu\right)$ behavior significantly differs. The parametric region to the left of the boundary corresponds to the trivial phase of the SC wire. Here the conductance exhibits a set of individual resonances induced by the zero-energy ABSs which are localized in one of the arms. In general, the map has the layered texture due to the two-channel interference and the dependence of coupling parameters $\tau_{i\sigma}$ on $h$ and $\mu$. 

To the right of the boundary (or inside the parabola), the S-segment is in the nontrivial phase. As was discussed before, here the $h$-dependence of conductance demonstrates the plateau and increasing oscillations as the Zeeman energy rises. Consequently, it results in the appearance of solid lines (in contrast to the trivial area outside the parabola) at $\mu<0.5$ where $G$ is close to $4G_{0}$. These lines of maxima are separated by intervals of minimum conductance. The latter arise since, according to the approximate solution \eqref{G_topo0}, the Bogolyubov coefficients are again the functions of $h$ and $\mu$ leading to the possibility of transmission antiresonances. In turn, such minima have to disappear if $\phi=\pi/2$ which is confirmed in Fig. \ref{5}b, where $G=4G_{0}$ while $A\approx0$. The trivial conductance in this situation is only slightly modified. As the Zeeman energy grows, the layered pattern on the map is restored. Comparing Figs. \ref{5}a and \ref{5}b at the high $h$, it can be seen that the conductance enhancement at $\phi=\pi/2$ occasionally occurs both in the case of the bulk ABS with zero energy and in the case when the first two excitations have energies of the same order.

When the S-section is in the nontrivial phase and $\mu>0.5$ the resonant lines in Fig. \ref{5}a and the resonant region in Fig. \ref{5}b are suppressed. A formal reason is shown in Fig. \ref{5}c where the MCs probability densities are represented. One can see that the second MC of the lowest-energy excitation is localized at the S/N2 interface and exponentially decays into the bulk of the bottom arm becoming disconnected from the normal contact via this path (the effect of its weak connection via the top arm on the conductance will be discussed below). In fact, we have transport in the single Andreev dot \cite{vuik-19} and the absence of the AB effect. Indeed, the area at $\mu>0.5$ acquires no changes for different $\phi$.

To reveal the physics of the feature shown in Fig. \ref{5}c it must be taken into account that the spectrum of the isolated homogeneous N2-subsystem defined by the Hamiltonian $\hat{H}_{n=2}$ with $V_{2i}=\varepsilon_{2}$, $h_{2i}=0$, $\Delta_{2i}=0$ (see expression \eqref{Hn}) is given by
\begin{equation} \label{N2_spectrum} 
E_{l}=\varepsilon_{2}-\mu-t\cos\frac{\pi l}{N_{2}+1}\pm \alpha_{2}\sin\frac{\pi l}{N_{2}+1},
\end{equation}
where $l=1,...,N_{2}$. Therefore, for the chosen parameters the largest positive value of the chemical potential corresponding to the zero-energy eigenstate is $\mu\approx0.5$. Next, let us consider the situation when the lowest-energy excitation of the N1+S-subsystem interacts with the one of the N2-subsystem (the hybridization is described by a parameter $T$). If the SC wire is in the topologically nontrivial phase and $\mu>0.5$, then $E_{N2} \gg E_{N1+S}$. Additionally assuming for simplicity that the coupling between the subsystems is weak, i.e. $E_{N2} \gg T$, the eigenvalues and eigenstates in such a model are $E_{1}\approx E_{N1+S}$, $E_{2}\approx E_{N2}$ and $\psi_{1}\approx\left[1,0\right]^{T}$, $\psi_{2}\approx\left[0,1\right]^{T}$, respectively. Thus, if $\mu>0.5$ the device lowest-energy state tends to settle in the N1+S-section. In other words, a local modification of the electrostatic potential in one of the normal regions (in practice by means of gate electrodes) can control the AB effect, allowing or preventing the MC leakage from the superconducting wire.

\begin{figure}[!htb]
	\begin{center}
		\includegraphics[width=0.525\textwidth]{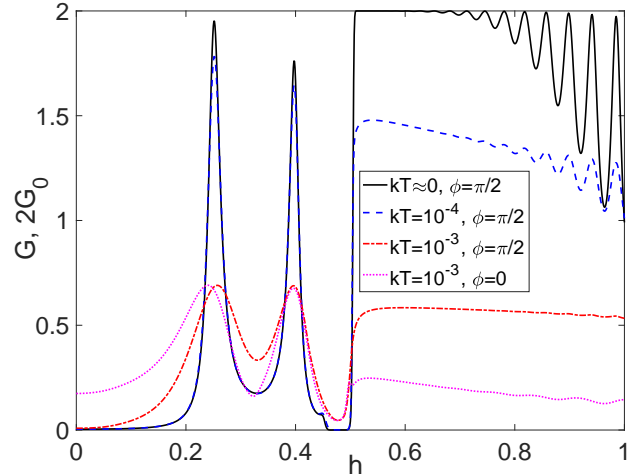}
		\caption{\label{6} Influence of nonzero temperature on the conductance for the 'steep-steep' type of the $V\left(j\right),~\Delta\left(j\right)$ profiles.}
	\end{center}
\end{figure}

Interestingly, the $G\left(h>h_{c1}\right)$ behavior at $\mu>0.5$ shown in Fig. \ref{5}d means that the second MC may not be completely isolated from the normal contact affecting the transport through the upper arm. Its influence is negligible immediately after the topological phase transition in the SC wire leading to the $2G_{0}$-quantized peaks due to the resonant Andreev reflection mediated by the first Majorana mode (see solid curve at $h\approx0.4$). The wave function $\psi_{2}^{M}$ penetrates more into the N1+S-part of the device as the Zeeman energy increases resulting in the direct coupling of the second MC with the reservoir. According to Vuik et al. (see the expression (16) and its discussion in \cite{vuik-19}), the two-channel interference is able to induce resonances with the amplitude $2G_{0}<G\leq4G_{0}$ and antiresonances $G=0$. Both are clearly seen in Fig. \ref{5}d. If the S-section is elongated, then the effect of the second MC is reduced (see dashed curve). In particular, the $2G_{0}$-maxima transform to the plateau and the subsequent resonances have a smaller amplitude than that of the shorter SC wire.  

\subsection{\label{sec4.4} Disorder and temperature factors}

\begin{figure*}[!tb]
	\begin{center}
		\includegraphics[width=1.05\textwidth]{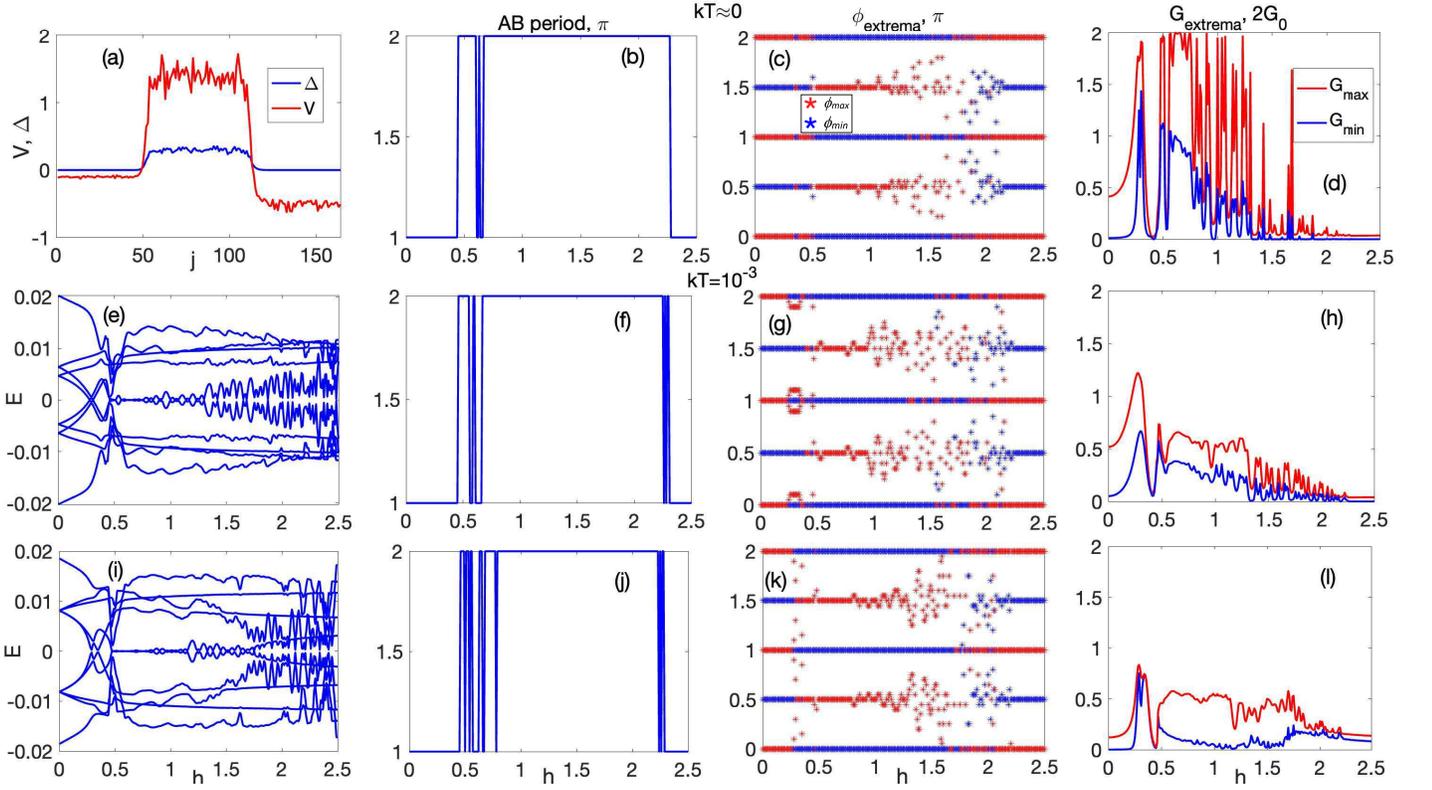}
		\caption{\label{AB_set2} Influence of random disorder and temperature factor on the properties of the low-energy Aharonov-Bohm effect. (a) Profiles of electrostatic and superconducting pairing potentials with incorporated single disorder realization for the 'steep-steep' case; (e,i) $h$-dependences of excitation spectrum; (b,f,j) $h$-dependences of Aharonov-Bohm period; (c,g,k) $h$-dependences of Aharonov-Bohm phase corresponding to the conductance maxima (red stars) and minima (blue stars); (d,h,l) maximum (red) and minimum (blue) conductances as functions of Zeeman energy. $kT\approx0$ in (b-d); $kT=10^{-3}$ in (f-h) and (j-l). The top and middle rows of the plots correspond to the same disorder realization displayed in (a). Parameters: $\sigma_{1}=2, \sigma_{2}=3$, $\sigma_{\varepsilon}=0.1\varepsilon$, $\sigma_{\Delta}=0.1\Delta$.}
	\end{center}
\end{figure*}

At the end of the original part, let us discuss the influence of the factors which are obviously encountered in practice, disorder and finite temperature, on the established transport features. In Fig. \ref{6} the effect of nonzero temperature on the conductance of the clean device is displayed for the 'steep-steep' configuration of the $V,~\Delta$ profiles. The resonances induced by the trivial zero-energy ABSs begin to broaden at $kT\sim10^{-4}$ which is of the same order of magnitude as the broadening parameters of the effective model, $\Gamma_{ij},~\Delta\Gamma_{ij},~\Delta\overline{\Gamma}_{ij}$. In turn, at $h>h_{c1}$ the plateau height decreases and the conductance oscillations are smoothed out. Indeed, such a modification is predictable since at low bias voltages the conductance is a convolution of transmission coefficient with the Fermi function derivative \cite{bruus-04}. As shown by the dot-dashed curve the separation between the two trivial maxima and the plateau related to the nontrivial phase persists at the experimentally available temperatures, $T\sim10$ mK. Moreover, the flux amplification effect discussed above is still present. Actually, the maximal conductance value increases by more than two times when the AB phase changes from $\phi=0$ to $\phi=\pi/2$ (compare dotted and dot-dashed curves). In this case, the height of the peaks in the trivial phase does not change.

In Fig. \ref{AB_set2} the influence of both random disorder and nonzero temperature on the properties of the AB effect is displayed. To investigate the former factor we have inserted fluctuations of the electrostatic and SC pairing potential in the device Hamiltonian \eqref{HD}. These terms are modeled by uncorrelated Gaussian distributions with zero means and standard deviations $\sigma_{\varepsilon,\Delta}$, i.e. $V_{dis}\sim N\left(0, \sigma_{\varepsilon}\right)$, $\Delta_{dis}\sim N\left(0, \sigma_{\Delta}\right)$. The $V\left(j\right)$ and $\Delta\left(j\right)$ profiles with the incorporated single disorder realization are plotted in Fig. \ref{AB_set2}a for $\sigma_{\Delta}=0.1\Delta$, $\sigma_{\varepsilon}^{m}=0.1\varepsilon_{m}$ (where $m=1,2,S$). In general, the spectrum calculations for the different disorder strengths and realizations repeat the conclusions obtained in earlier studies of the individual SC wire (see, e.g., \cite{sau-13}). In particular, as long as the deviation in the S-section $\sigma_{\varepsilon}^{S}$ is less than the topological gap, the nontrivial phase in our inhomogeneous device survives, i.e. the separated MCs of the lowest-energy excitation, which exponentially decay into the SC bulk, can be observed. Since the increase of $h$ leads to the topological gap suppression (at the high magnetic fields the p-wave pairing amplitude $\sim \alpha_{S}\Delta/h$), the gap in the excitation spectrum shown in Fig. \ref{AB_set2}e actually collapses at $h>2$. However, at the weak and moderate fields ($0.6<h<1$) the large enough gap still persists. The analysis of the MCs spatial distributions in the disordered structure reveals the existence of all three types of the states with energies close to zero (inhomogeneous ABSs, MBSs and bulk ABSs).
	
This is also backed by the AB effect features which are consistent with the results of Table \ref{tab1} at both zero and nonzero temperatures. The inhomogeneous ABSs and MBSs can be distinguished by the AB period. In the latter situation, it oscillates and then stabilizes at $2\pi$ as the Zeeman energy increases, while in the former, the AB period is always equal to $\pi$ (Fig. \ref{AB_set2}b). Interestingly, an additional proof of the disorder-induced decrease of the topological gap is that the region of $h$ where the AB period oscillates, indicating the MBS-mediated transport, now appears at the lower fields $h\approx0.6$ leaving no room for the region where the AB effect is constantly $\pi$-periodic due to $A\approx0$ (compare Figs. \ref{AB_set2}b and \ref{AB_set1}b). In turn, the MBS and bulk ABS cases can be separated taking into account both the AB period and robust positions of the conductance extrema $G\left(\phi\right)$. For the bulk-type excitations, the period does not change and the sustained AB phases are $\pi n$. It is clearly seen in Fig. \ref{AB_set2}c that the other extrema phases are randomly distributed at $1<h<2$. The observed sustainability of the results is explained by the fact that weak disorder does not change the spatial behavior of the states providing the low-energy interference transport. Comparison of Figs. \ref{AB_set2}b,c with  Figs. \ref{AB_set2}f,g shows that finite temperature has no detrimental effect on the properties of the AB oscillations. 

In the bottom row of Fig. \ref{AB_set2} the same dependencies as in the middle row are displayed but for the different disorder realization. It can be seen that the modification of the AB effect characteristics in Figs. \ref{AB_set2}j,k is only quantitative. Importantly, before the topological phase transition ($h<h_{c1}$) the possibility of conductance amplification by the magnetic flux depends on the specific view of the potentials profiles. In Fig. \ref{AB_set2}e there are two almost degenerate ABSs with close-to-zero energy. Returning to the effective model, it means $A,~B\ll\Gamma_{i\sigma,j\sigma'}$ leading to the stronger AB-phase dependence of the conductance according to \eqref{G_triv0}. In Fig. \ref{AB_set2}l, $B\gg A,~\Gamma_{i\sigma,j\sigma'}$ around the conductance resonances and the AB oscillations amplitude is much smaller. At $h>h_{c1}$ the magnetic-flux enhancement persists over a wide range of the Zeeman energies as long as the gap in the excitation spectrum is significant.

\section{\label{sec5}Summary}

In the reported study we have analyzed the features of coherent local transport associated with the presence of Andreev and Majorana states in a nonuniform one-dimensional N/S/N structure. Unlike traditional measurements used to probe MBSs, when the contact interacts with one end (or two contacts are connected to the opposite ends), the proposed curved geometry additionally allows us to study interference transport. The inhomogeneity in the device has been modeled by the profiles of electrostatic and SC pairing potentials smoothly changing in the vicinity of the two N/S boundaries. 

It has been demonstrated that even if the SC wire is in the trivial phase Andreev reflection in such a structure leads to the appearance of inhomogeneous ABSs which are localized in the different normal arms and adjacent edges of the S-segment and can have zero energy. Taking it into account we have formulated the effective model that describes low-energy interference transport in the two noninteracting Andreev levels coupled in parallel with electron and hole reservoirs. Based on the scattering matrix formalism, the linear-response conductance in the low-temperature limit has been analyzed analytically including various limiting cases of inhomogeneous ABSs, bulk ABS and MBS.

In particular, we have focused on the consequences of the existing spatial nonuniformity. Depending on the inhomogeneity strength, the wave functions of the Majorana components, which constitute the trivial ABSs, behave differently. If the profiles change steeply at the given interface then the spatial distributions of both MCs are delocalized throughout the nearest N-section and overlap sizably resulting in the inhomogeneous zero-energy ABS. In turn, if the profile change is gentle the MCs overlap becomes comparable with the one exhibited by the topological MBS. As a result, the trivial quasi-MBS emerges. It has been shown that the width of conductance resonances associated with the different trivial excitations differs significantly. On the contrary, the resonances induced by the topological MBS are immune to changes in the smoothness degree.

We have demonstrated that if the SC wire is in the trivial case, then the AB effect is based on the interference processes involving two inhomogeneous ABSs localized in the opposite arms. In the nontrivial phase, the conductance oscillations can be determined by Andreev reflection on the Majorana modes composing the single lowest-energy excitation, either MBS (at the low magnetic fields) or bulk ABS (at the high magnetic fields). Then, in order to effectively distinguish between these three types of low-energy excitations, it is necessary to analyze the features of both AB period and extrema of the conductance oscillations which do not require fine tuning of the parameters.

Essentially, if the conductance peak is induced by the topological MBS with exponentially small energy, its height can be always increased to the quantized value by the magnetic flux. The maximum is equal to $4G_0$ at $\phi=\pi (n+1/2)$ since the AB effect is determined by the transport only into the Majorana modes of the first excitation which are localized in the opposite arms of the device. In the case of the inhomogeneous and bulk ABSs, the interaction between the contact and separate arm is implemented via the MCs of both types leading to the different possible scenarios when the AB phase varies. Here the quantized conductance at $\phi=\pi (n+1/2)$ is achieved only accidentally with a proper choice of the tunneling coefficients. It has been demonstrated that the dependence of the AB effect properties on the ABS type is retained in the presence of weak disorder and at temperatures available in modern experiments.

\begin{acknowledgments}
We thank A. D. Fedoseev and M. S. Shustin for fruitful discussions. The study was supported by Russian Science Foundation, Project No. 19-12-00167.

\end{acknowledgments}

\bibliography{Majorana}

\end{document}